\documentclass[twocolumn]{aastex631}
\usepackage{amsmath,epsfig,url}
\usepackage{graphicx,epstopdf,float,color,array}
\usepackage{rotating,multirow}

\shorttitle{MOSDEF SOM}
\shortauthors{Sanjaripour et al.}

\begin{document}
\title{\textbf{Application of Manifold Learning to Selection of Different Galaxy Populations and Scaling Relation Analysis}}
\author[0009-0009-3048-9090]{Sogol Sanjaripour}
\affiliation{Department of Physics and Astronomy, University of California Riverside, Riverside, CA 92521, USA}
\author[0000-0003-2226-5395]{Shoubaneh Hemmati}
\affiliation{IPAC, California Institute of Technology, Pasadena, CA 91125, USA}
\author[0000-0001-5846-4404]{Bahram Mobasher}
\affiliation{Department of Physics and Astronomy, University of California Riverside, Riverside, CA 92521, USA}
\author[0000-0003-4693-6157]{Gabriela Canalizo}
\affiliation{Department of Physics and Astronomy, University of California Riverside, Riverside, CA 92521, USA}
\author[0000-0001-6386-7371]{Barry Barish}
\affiliation{LIGO Laboratory, California Institute of Technology, Pasadena, CA 91125, USA}
\affiliation{Department of Physics and Astronomy, University of California Riverside, Riverside, CA 92521, USA}
\author[0000-0003-4702-7561]{Irene Shivaei}
\affiliation{Centro de Astrobiología (CAB), CSIC-INTA, Madrid, 28850, Spain}
\author[0000-0002-2583-5894]{Alison L. Coil}
\affiliation{Center for Astrophysics and Space Sciences, Department of Physics, University of California, San Diego, CA 92093-0424, USA}
\author[0000-0003-3691-937X]{Nima Chartab}
\affiliation{IPAC, California Institute of Technology, Pasadena, CA 91125, USA}
\author[0000-0001-8019-6661]{Marziye Jafariyazani}
\affiliation{IPAC, California Institute of Technology, Pasadena, CA 91125, USA}
\author[0000-0001-9687-4973]{Naveen A. Reddy}
\affiliation{Department of Physics and Astronomy, University of California Riverside, Riverside, CA 92521, USA}
\author[0000-0001-6004-9728]{Mojegan Azadi}
\affiliation{Center for Astrophysics | Harvard \& Smithsonian, Cambridge, MA 02138, USA}

\email{sogol.sanjaripour@email.ucr.edu}

\journalinfo{\textcopyright\ 2024. All rights reserved. To be submitted to the Astrophysical Journal.}

\begin{abstract}
The growing volume of data produced by large astronomical surveys necessitates the development of efficient analysis techniques capable of effectively managing high-dimensional datasets. This study addresses this need by demonstrating some applications of manifold learning and dimensionality reduction techniques, specifically the Self-Organizing Map (SOM), on the optical+NIR SED space of galaxies, with a focus on sample comparison, selection biases, and predictive power using a small subset. To this end, we utilize a large photometric sample from the five CANDELS fields and a subset with spectroscopic measurements from the KECK MOSDEF survey in two redshift bins at $z\sim1.5$ and $z\sim2.2$. We trained SOM with the photometric data and mapped the spectroscopic data onto it as our study case. We found that MOSDEF targets do not cover all SED shapes existing in the SOM. Our findings reveal that Active Galactic Nuclei (AGN) within the MOSDEF sample are mapped onto the more massive regions of the SOM, confirming previous studies and known selection biases towards higher-mass, less dusty galaxies. Furthermore, SOM were utilized to map measured spectroscopic features, examining the relationship between metallicity variations and galaxy mass. Our analysis confirmed that more massive galaxies exhibit lower [O{\sc iii}]/H$\beta$ and [O{\sc iii}]/[O{\sc ii}] ratios and higher H$\alpha$/H$\beta$ ratios, consistent with the known mass-metallicity relation. These findings highlight the effectiveness of SOM in analyzing and visualizing complex, multi-dimensional datasets, emphasizing their potential in data-driven astronomical studies.
\end{abstract}


\section{Introduction}
The field of observational astronomy has undergone transformational changes in recent years, largely due to the commissioning of new imaging detectors, multi-slit or fiber-fed spectrographs, and new observatories both on the ground and in space. These advancements have significantly increased the volume of available data from various galaxy surveys. Extensive imaging and spectroscopic initiatives such as the Sloan Digital Sky Survey (SDSS; \citealt{York2000}), the Cosmic Evolution Survey (COSMOS; \citealt{Scoville2007}), and the Cosmic Assembly Near-infrared Deep Extragalactic Legacy Survey (CANDELS; \citealt{Grogin2011}; \citealt{Koekemoer2011}) have collected large amounts of data on galaxies over time. The data volume will further increase with forthcoming contributions from the Dark Energy Spectroscopic Instrument (DESI; \citealt{Abareshi2022}), the Prime-Focus Spectrograph (PFS) on the 8m Subaru Telescope (\citealt{Tamura2016}), the Euclid Space Telescope (\citealt{Racca2016}), the Rubin Observatory Legacy Survey of Space and Time (LSST; \citealt{LSST2009,Rubin2022}), the Roman Space Telescope (\citealt{Domber2022}), and SPHEREx (\citealt{Doré2015}).

\begin{figure*}[htbp]
\centering
  \includegraphics[trim=0cm 0cm 0cm 0cm, clip,width=0.98 \textwidth] {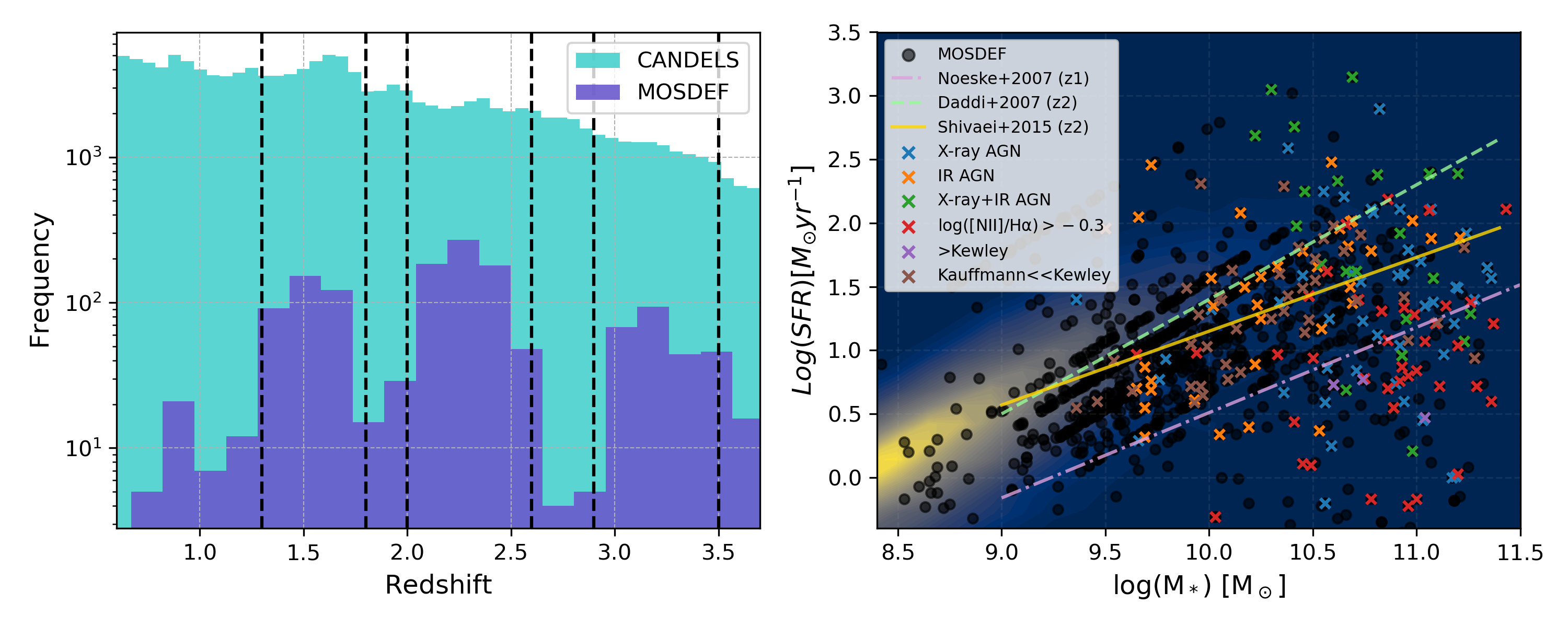}
\caption{\textbf{Left:} The redshift distribution of the MOSDEF spectroscopic sample on top of the underlying photometric data from the five CANDELS fields. The black dashed lines show the three redshift bins asserted by the MOSDEF team (\citealt{Kriek2015}). \textbf{Right:} Distribution of the sample on the star formation rate vs. stellar mass plane. The underlying 2d histogram shows CANDELS galaxies in the redshift range of $0.6<z<3.6$ to match that of MOSDEF. Black circles are the MOSDEF targets and the colored crosses are the identified AGNs with different techniques. The light green, purple, and gold dashed lines show the main sequence of starforming galaxies at $z = 1,2, 2.5$ respectively from \citealt{Daddi2007,Noeske2007} and \citealt{Shivaei2015}.}
\label{fig:sample}
\end{figure*}


Extracting optimal scientific information from these extensive datasets, whether single or combined, is unachievable without the development of new and innovative techniques. In this context, Machine Learning and Deep Learning (ML/DL) have emerged as indispensable tools for processing large datasets (e.g., \citealt{Ball2010, Acquaviva2015, baron2019, Huertas2023}). ML/DL applications in astronomy encompass a wide range of tasks, including the identification and classification of targets (e.g., \citealt{stargalaxy2017, merger2018, Lensfind2018, Faisst2019}), measuring physical properties of galaxies from observed (e.g., \citealt{Collister2004, Kind2013, Hemmati2019b, Davidzon2019, Surana2020}) or simulated data (e.g., \citealt{Raouf2016, Lovell2019, Davidzon2019, Laigle2019}), and conducting hydrodynamical cosmological simulations (e.g., \citealt{Villaescusa_Navarro2021, Villaescusa_Navarro2022}). One particularly notable application of ML/DL in astronomy is learning the high-dimensional manifold of galaxy data. Techniques like Self-Organizing Maps (SOM; \citealt{Kohonen1982}) or Uniform Manifold Approximation and Projection (UMAP; \citealt{McInnes2018}) can map these multi-dimensional data to more manageable dimensions, revealing hidden trends and correlations. Over the last few years, these techniques have gained significant attention and have been increasingly adopted in various studies (e.g., \citealt{Geach2012, Davidzon2022, Busch2022,Chartab2023, Torre2024}).

\citet{Geach2012} introduced Self-Organizing Maps (SOMs) for visualizing, exploring, and mining large astronomical survey catalogs, demonstrating their utility in object classification and photometric redshift estimation. Subsequent work extended these efforts for weak-lensing cosmology calibration by mapping spectroscopic subsamples to reduced-dimension grids trained on extensive photometric datasets (e.g., \citealt{Masters2015, Masters2019, Hemmati2019, McCullough2024}). Although the primary focus of these studies has been on photometric redshift calibration, they illustrate the potential for comparing and combining datasets acquired through different methods. This kind of domain transfer or matching has also been utilized to measure the physical properties of galaxies. For example, photometric observations have been aligned with stellar population synthesis models to measure stellar masses (e.g., \citealt{Hemmati2019b}), hydrodynamical simulations (e.g., \citealt{Davidzon2019}), and observed data with extended wavelength coverage, such as Spitzer MIPS data, for more precise measurements of star formation rates (e.g., \citealt{Davidzon2022}). Additionally, spectroscopic subsamples have been mapped to photometric SOMs out to $z\sim 1.5$ to measure features such as the strength of 4000 angestrom break (D4000) (e.g., \citealt{Jafariyazani2024}).

In this study, we extend previous efforts by analyzing spectroscopic data from the Keck Multi-Object Spectrometer for Infra-Red Exploration (MOSFIRE; \citealt{McLean2012}) Deep Evolution Field (MOSDEF; \citealt{Kriek2015}) survey, which provides emission line indices for galaxies within the redshift range $1.3 < z < 3.5$. This spectral coverage facilitates the measurement of a wide array of spectral features. The MOSDEF galaxies lie in the Cosmic Assembly Near-infrared Deep Extragalactic Legacy Survey (CANDELS; \citealt{Koekemoer2011, Grogin2011}) fields, which offer very deep photometric data. We learn the galaxy SED manifold from this photometric data using a consistent set of broadband filters and map the high-quality MOSDEF spectral measurements onto the reduced-dimension SOM. Our objectives include exploring sample comparisons and biases in selections, uncovering correlations between observed properties, and demonstrating the predictive power using a limited subset of data.

We specifically investigate the MOSDEF sample and the active galactic nuclei (AGNs) within it to highlight potential selection biases related to SED types. By examining AGN subsamples within MOSDEF, we assess whether AGNs are preferentially found in specific types of galaxies. Considering the high cost and time required for spectroscopic observations, the ability to use a subset of this data to infer properties for similar photometric SEDs would be highly advantageous.

The paper is structured as follows: Section \S 2 details the datasets used for training and testing. Section \S 3 discusses the methodology. Section \S 4 presents the results, and Section \S 5 considers the future implications and applications of this approach within the context of larger surveys. Throughout this work all magnitudes are given in the AB system (\citealt{Oke1983}).

\section{Sample and Data}

\begin{figure*}[htbp]
\centering
  \includegraphics[trim=0cm 1cm 0cm 0cm, clip,width=0.98 \textwidth] {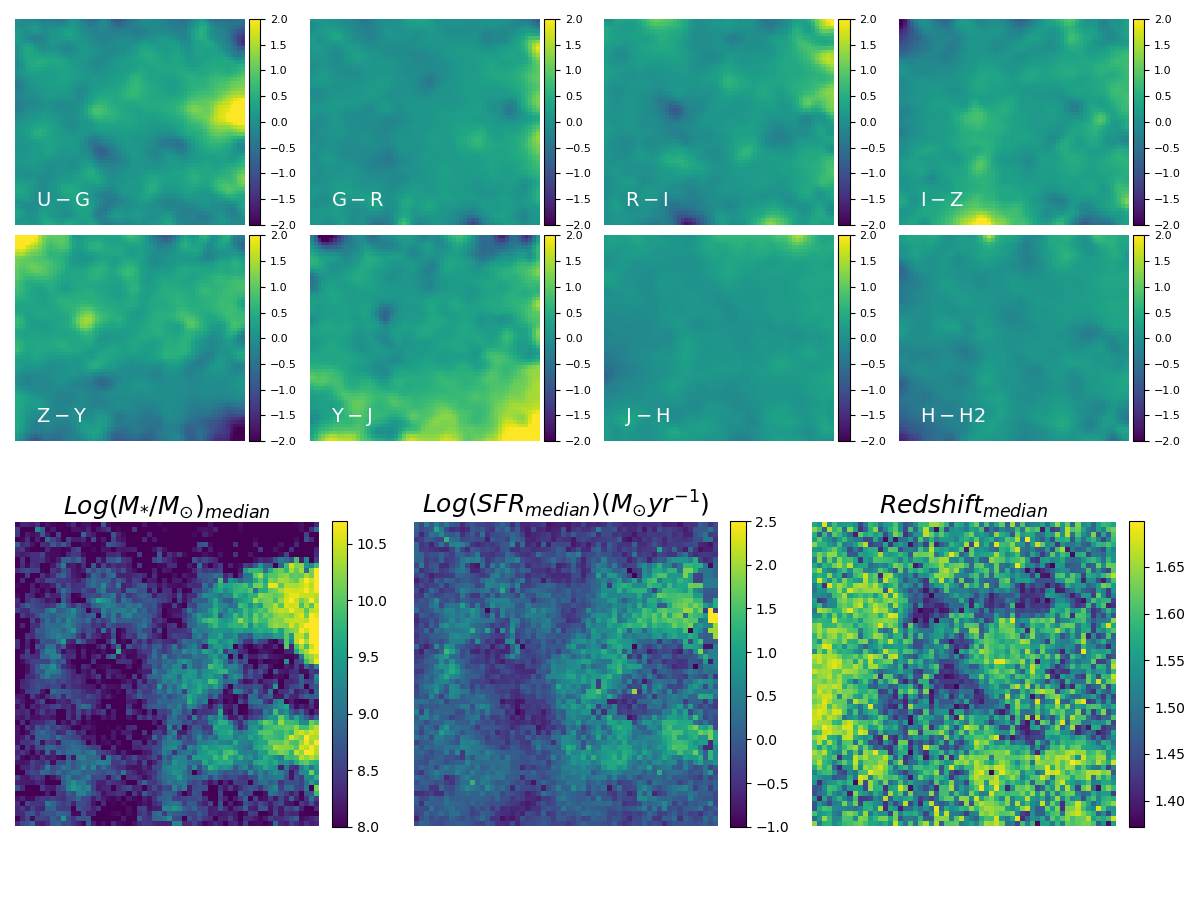}
\caption{ 
\textbf{Top two rows} represent the component maps of the trained SOM at $1.37 < z < 1.70$, resulting from learning the 8 dimensional color manifold. The color bars represent the respective colors for individual SOMs. In the \textbf{bottom row} the SOM cells are color-coded based on the median of normalized stellar mass, normalized Star Formation Rate (SFR), and galaxy redshift of CANDELS galaxies used in the training process, respectively from left to right. These parameters are derived from the SEDs of galaxies, involving all nine colors (8 dimensions)}
\label{fig:somtrain1}
\end{figure*}

In this section, we provide a summary of the photometric data from the CANDELS survey and the spectroscopic data from the MOSDEF survey. For comprehensive details on each dataset, we refer readers to the extensive literature available on these surveys. While we have chosen the MOSDEF sample as a representative example, the methodologies and analyses presented in this work can be readily extended to other samples.

\subsection{CANDELS photometric data}

CANDELS is a deep imaging survey of five distinct fields in the sky, conducted using the WFC3/IR and ACS cameras on board the Hubble Space Telescope (HST). The HST observations are homogeneously combined with a wealth of ancillary space- and ground-based data, spanning wavelengths from the ultraviolet to the mid-infrared.  The CANDELS survey spans an aggregate area of approximately $\sim0.2$ square degrees and the WFC3 F160W observations achieve a limiting magnitude depth of approximately 26.5 at a 5$\sigma$ significance level. All source detections in the CANDELS fields were performed in the WFC3 F160W band (at 1.6 $\mu$m), and photometry was generated using the Template FITting algorithm (TFIT; \citealt{Laidler2007}).An overview of the observations, data reduction, and measurement of the physical parameters of galaxies in the five CANDELS fields is presented in the following studies: the GOODS-South field \citep{Guo2013}, the UDS field \citep{Galametz2013}, the COSMOS field \citep{Nayyeri2017}, the Extended Groth Strip field \citep{Stefanon2017}, and the GOODS-North field \citep{Barro2019}.

To utilize targets from all CANDELS fields with observations in non-unified filter sets for learning the manifold of galaxy SEDs, we follow \cite{Hemmati2019}, who transferred the photometric measurements from the original CANDELS filters into a 9-band Rubin/LSST+ROMAN filter set ($u_{\lambda 367 nm}$, $g_{\lambda 483 nm}$, $r_{\lambda 622 nm}$, $i_{\lambda 755 nm}$, $z_{\lambda 869 nm}$, $y_{\lambda 1060 nm}$, $J_{\lambda 1290 nm}$, $H_{\lambda 1580 nm}$, $H2_{\lambda 2130 nm}$) by linearly interpolating between the deepest straddling CANDELS bands. We also applied a magnitude cut, selecting objects with magnitudes between 15 and 29 in all bands. The deliberate choice of filter set is justified as it covers the same wavelength range as the CANDELS observations and facilitates future comparisons with all-sky data from LSST+ROMAN. 

\begin{figure*}[htbp]
\centering
  \includegraphics[trim=0cm 1cm 0cm 0cm, clip,width=0.98 \textwidth] {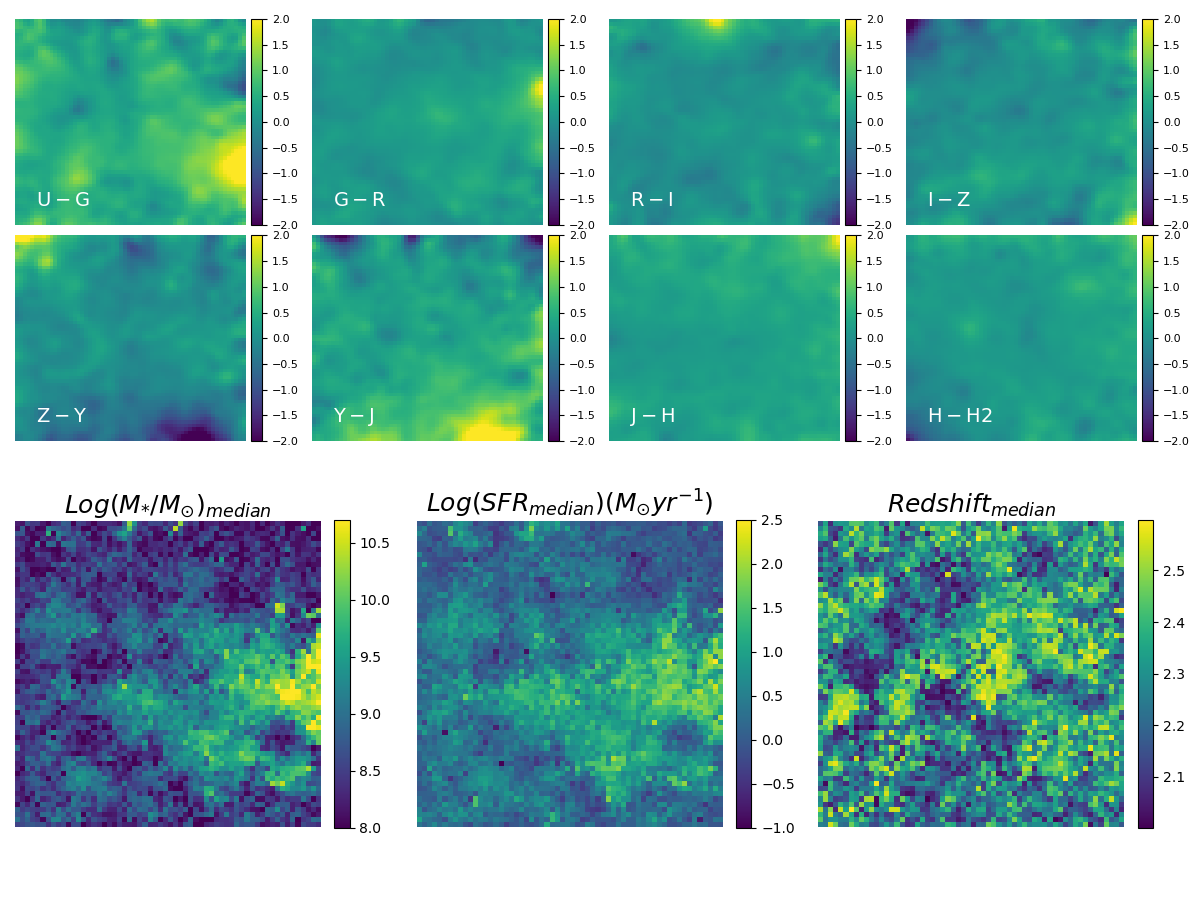}
\caption{Similar to Figure \ref{fig:somtrain1}, trained on the $2.0<z<2.6$ galaxies.}
\label{fig:somtrain2}
\end{figure*}

\subsection{The MOSDEF sample}

With 48.5 nights of MOSFIRE observations on the Keck I Telescope, the MOSDEF survey obtained near-infrared spectra in the J, H, and K bands for over 1,500 CANDELS galaxies within the redshift range $1.37 < z < 3.80$ (\citealt{Kriek2015}). They strategically focused on galaxies within three distinct redshift ranges: $1.37 < z < 1.70$, $2.09 < z < 2.61$, and $2.95 < z < 3.80$ (see Figure \ref{fig:sample}). This selection was guided to exploit atmospheric transmission windows, allowing for measurements of diagnostic rest-frame optical emission lines such as [O{\sc ii}] $\lambda\lambda 3727, 3729$, H$\beta$, [O{\sc iii}]$\lambda\lambda 4960, 5008$, H$\alpha$, [N{\sc ii}]$\lambda\lambda 6550, 6585$, and [S{\sc ii}]$\lambda\lambda 6718, 6732$. To enhance the quality of the data, MOSDEF employed a dithering technique, which helped to avoid potential artifacts and increased the signal-to-noise ratio of their acquired spectra.

The MOSDEF survey’s large sample size and robust measurements of key emission lines make it particularly valuable and suitable for our study. Figure \ref{fig:sample} presents the MOSDEF galaxies overlaid on the general CANDELS population, depicting both the redshift distribution and the star formation vs. stellar mass plane. We utilized the measured spectroscopic redshifts, stellar masses, and star formation rates from the published catalogs (\citealt{Reddy2015}, \citealt{Kriek2015}). The plot also includes the main sequence of star formation at $z\sim 1, 2, 2.5$ from the literature (\citealt{Noeske2007, Daddi2007, Shivaei2015}), providing a visual guide for assessing potential biases in target selection towards more or less massive, star-forming objects or AGNs. Unlike the green and purple lines, the yellow dashed line from \cite{Shivaei2015} is derived from the same MOSDEF galaxies in the second redshift bin, with the SFR based on dust-corrected $H\alpha$. The detection of H$\alpha$ and H$\beta$ in a large sample within this redshift range enabled accurate dust correction and, consequently, precise star formation rate measurements for star-forming galaxies.

In addition to the MOSDEF sample as a whole, the survey includes the identification of Type 2 AGNs using X-ray, infrared, and optical measurements (represented by colored crosses in the right panel of Figure \ref{fig:sample}, \citealt{Azadi2017}). This allows for detailed subsample comparisons, enabling us to compare different selection techniques using a holistic view of the entire SED, as explained in the next sections. When studying selection biases, it is more effective to adopt this holistic approach rather than examining only the redshift distribution, a single magnitude, or two measured properties such as mass and star formation rate (SFR).

\section{Manifold learning of galaxy SEDs}

\begin{figure*}[htbp]
\centering
  \includegraphics[trim=0cm 0cm 0cm 0cm, clip,width=0.99 \textwidth] {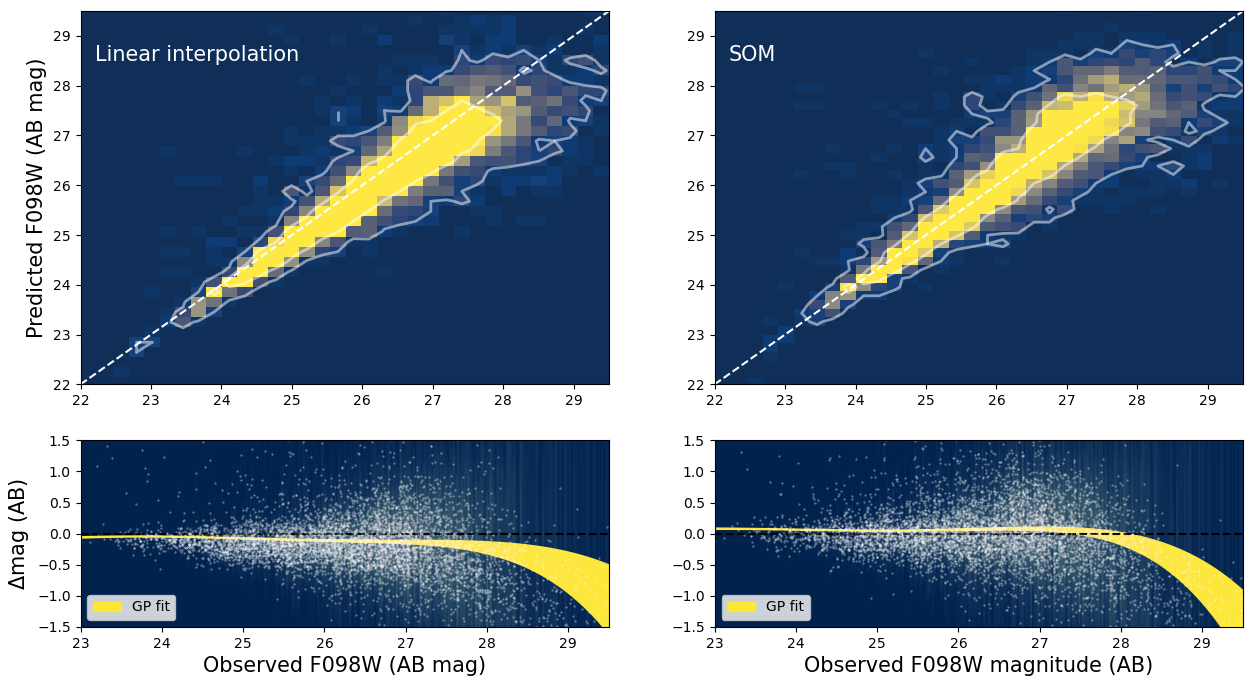}
\caption{Augmenting a missing band in the photometric catalog using \textbf{Left}: linear interpolation and \textbf{Right}: the SOM. The top panels show contours on the scatter plots of the predicted vs. real F098W magnitudes for a test subsample, and the bottom panels show a Gaussian regression fit to the delta magnitudes. Both methods predict the missing fluxes with reasonable accuracy comparable to photometric uncertainties.}
\label{fig:missingband}
\end{figure*}

SOMs are an effective technique for manifold learning, particularly useful for uncovering hidden patterns in high-dimensional astronomical data while preserving topological features in the reduced dimensions \citep{Kohonen1982}. Training SOMs involves iteratively adjusting the weight vectors of the grid units in an unsupervised manner to align them with the input data. In this context, training refers to the unsupervised learning process by which the SOM adapts to represent the structure of the input data without the need for labeled outputs. Each unit (or neuron) in the SOM grid possesses an associated weight vector, which serves as a prototype representing a position in the input data space.

The Best Matching Unit (BMU) is identified by determining the unit whose weight vector is closest to a given input data point, typically using a distance metric such as the Euclidean distance. During the training process, the weight vectors of the BMU and its neighboring units are updated to minimize their distances to the input data point, thereby adapting the map to reflect the underlying patterns in the data. The stochastic nature of SOMs, arising from the random initialization of weight vectors and the random order in which input data points are presented during training, can lead to varied outcomes across different training sessions. Parameters such as map size and learning rate, which determines how much the weight vectors are adjusted during training, can also result in varied outcomes. We used the \textit{SOMPY} package (\citealt{sompy}) to implement SOMs. Following \cite{Davidzon2019}, we selected a square SOM and determined through experimentation that a $60 \times 60$ grid size optimally balances detail resolution and computational efficiency, capturing dataset variations without compromising statistical significance.

The high-dimensional galaxy SED manifold we construct in this study is represented in an 8-dimensional color space derived from the 9-band photometries detailed in the previous section: u-g, g-r, r-i, i-z, z-Y, Y-J, J-H, and H-H2. We trained two separate SOMs for the two MOSDEF redshift bins, covering $1.3<z<1.7$ with 9,976 galaxies and $2.0<z<2.6$ with 14,130 galaxies. While the third redshift bin of MOSDEF could be similarly analyzed, it was excluded from this work due to lower AGN counts leading to insufficient statistics and the fact that not all diagnostic emission lines could be covered with MOSFIRE. Figures \ref{fig:somtrain1} and \ref{fig:somtrain2} show the trained SOM grid in the two redshift bins, respectively, where the top two rows in each are color-coded with the final 8 weights after training, and the bottom rows are color-coded based on the median values of stellar mass, SFR, and photometric redshift of the galaxies mapped to each cell directly from the CANDELS catalogs. 

Visual inspection of the SOMs color-coded by stellar mass and SFR reveals trends indicative of effective grouping of SED shapes. It is important to note that, although the SOM is trained on colors (i.e., normalized SEDs), a spatial trend with median stellar mass remains. This is expected, as stellar mass correlates tightly with brightness and also with SED shapes. Additionally, a strong correlation between the spatial trends in the stellar mass and SFR planes is observed; however, this partially arises from the interdependence of these measurements through SED fitting.

To determine the optimal configuration of our SOM, we used the quantization error, which measures the average distance between each data point and its closest neuron (BMU) in the SOM. A lower quantization error indicates a better representation of the input data. The chosen grid size of $60 \times 60$ is near the elbow point in the analysis of quantization error versus grid sizes, which signifies the optimal balance between error reduction and model complexity. This selection provided a satisfactory representation of the input data while minimizing the quantization error. Specifically, we achieved quantization errors of 0.84 and 0.77 for the two mentioned redshift ranges, respectively. The chosen grid size ensures that the SOM effectively captures the underlying structure of the data, facilitating accurate clustering and analysis of SED shapes.

\subsection{Augmenting Missing Fluxes}

Once SOMs are trained, the weights of the grid represent the input distribution, with similar data points clustering together and outliers tending to move towards the edges of the grid. Outliers in the photometric catalog, which may arise from non-detections or larger uncertainties in a band in addition to rare interesting physics, can skew the edge population of the SOM. To prevent the edges from being overpopulated with such non-detections, it is beneficial to address these issues in the catalog before learning and reducing the dimensions of the manifold.

In \S 2.1, we briefly discussed our linear interpolation strategy to transform all CANDELS fields into a unified set of filters, without addressing the non-detections in the original catalogs. \cite{Chartab2023} and \cite{Torre2024} employed manifold learning to fill gaps in photometric catalogs, quantifying the improvements in the resulting measured physical properties. Here, we tested two methods for augmenting missing fluxes before filter transformation: linear interpolation with the nearest neighbors and using the SOM itself to fill the gaps.

The simplest method for filling in missing data is linear interpolation, which does not assume any prior knowledge of the spectral energy distribution (SED). In contrast, using the SOM as a predictor for a missing band incorporates a data-driven prior derived from SEDs that show similarity in other bands. Despite the increased complexity, this approach leverages the inherent structure within the dataset to make more informed predictions. Both methods perform well, with the prediction uncertainties closely aligning with typical observed uncertainties in photometry. In Figure \ref{fig:missingband}, we present a test on the Y band (F098W), predicting F098W for a sample within the redshift range $2.0 < z < 2.6$ in the GOODS-S field, using linear interpolation (left panel) and the SOM (right panel). The Gaussian process regression on the delta magnitude of prediction versus observed values in the bottom panel shows excellent agreement up to ~28 AB mag (i.e., F098W $5\sigma$ completeness), with the SOM technique slightly outperforming. In this work, we filled in the gaps of the filters in use with a SOM trained on each field before filter transformation.

\begin{figure}[!htbp]
\centering
  \includegraphics[trim=3cm 0.5cm 0cm 0.5cm, clip,width=0.5\textwidth] {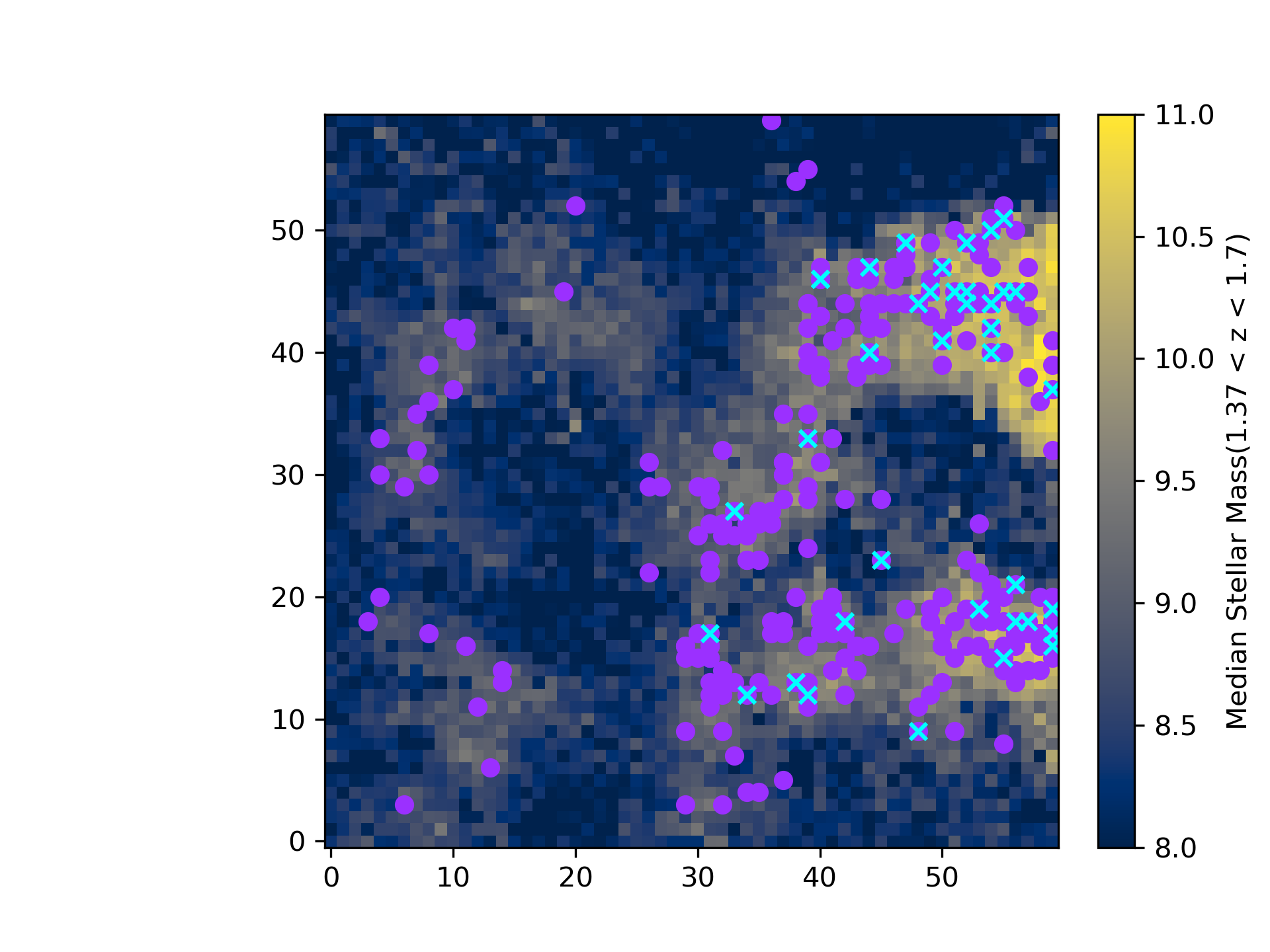}
  \includegraphics[trim=3cm 0.5cm 0cm 0.5cm, clip,width=0.5\textwidth] {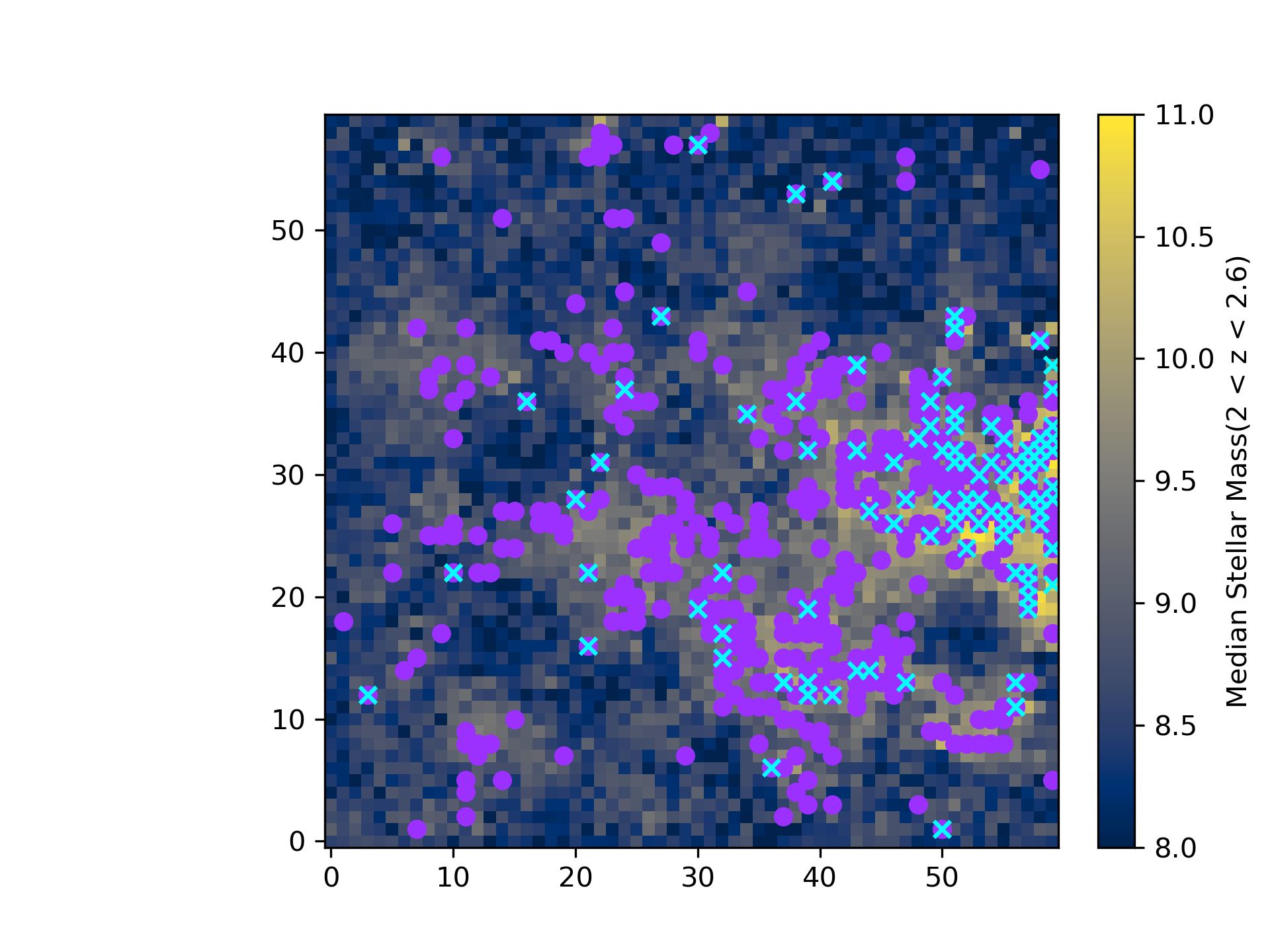}
\caption{MOSDEF galaxies mapped onto the trained SOM. \textbf{Top panel:} Displays the first redshift bin at $1.3 < z < 1.7$. \textbf{Bottom panel:} Depicts the redshift range $2.0 < z < 2.6$. The background color represents the median stellar mass of CANDELS galaxies within each redshift bin. Normal star-forming MOSDEF galaxies are indicated by purple circles, while MOSDEF AGNs are marked by cyan 'x' symbols.}
\label{fig:massdistribution}
\end{figure}


\section{Results}

In this section, we employ the reduced dimension projection of the learned manifold from the photometric SEDs to visualize the selection functions of the MOSDEF sample and the AGNs therein, thereby facilitating a more comprehensive sample comparison. Additionally, we present the scaling relations of the physical properties derived from the nebular emission lines within our learned 2D space.

\subsection{MOSDEF target selection}

Findings from various surveys and studies are sometimes extended, compared, or generalized to broader populations, yet the selection functions that shape these samples may not always be fully considered. Different selection criteria, such as mass-limited or magnitude-limited samples, or those with inherent incompleteness, can lead to assumptions that may not fully capture the diversity of the population at a given redshift. Conversely, some samples are treated as fully distinct, even when there are similarities or common members that are not immediately apparent, as is the case with Lyman break galaxies and Lyman alpha emitters. Understanding these selection effects is crucial for accurate interpretation and meaningful comparisons across studies. Mapping samples selected in different ways onto a unified reduced-dimensional manifold provides a straightforward method to visualize sample completeness, whether it occupies the full parameter space, as well as to measure the similarities or differences between them.

\begin{figure*}[htbp]
\centering
  \includegraphics[trim=5cm 4cm 4cm 5cm, clip,width=1 \textwidth] {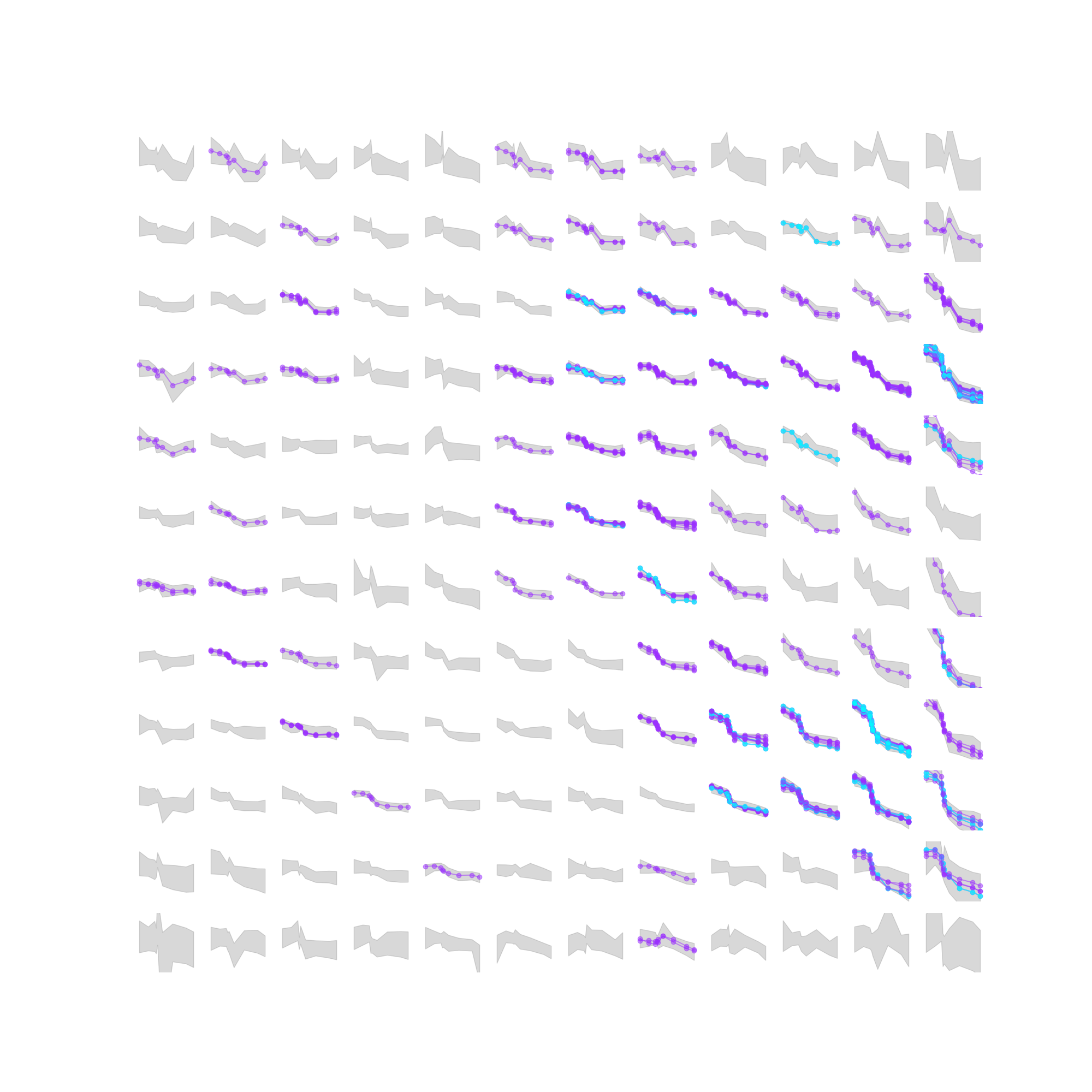}
\caption{The SED profiles of galaxies mapped onto the SOM with five times fewer cells. The gray SED curves represent the training set from the CANDELS survey, encompassing a variety of galaxy types, with the shading indicating the photometric errors. Overlaid in purple are the SEDs of normal star-forming galaxies from the MOSDEF survey, while the cyan SEDs correspond to AGNs from the same survey. All galaxies included in this analysis fall within the redshift range of $1.37 < z < 1.7$. As illustrated, SED shapes gradually change from cell to cell, with similar SEDs in neighboring cells and higher photometric errors at the edges of the SOM.}
\label{fig:SEDSOM1}
\end{figure*}

\begin{figure*}[htbp]
\centering
  \includegraphics[trim=5cm 4cm 4cm 5cm,width=1 \textwidth] {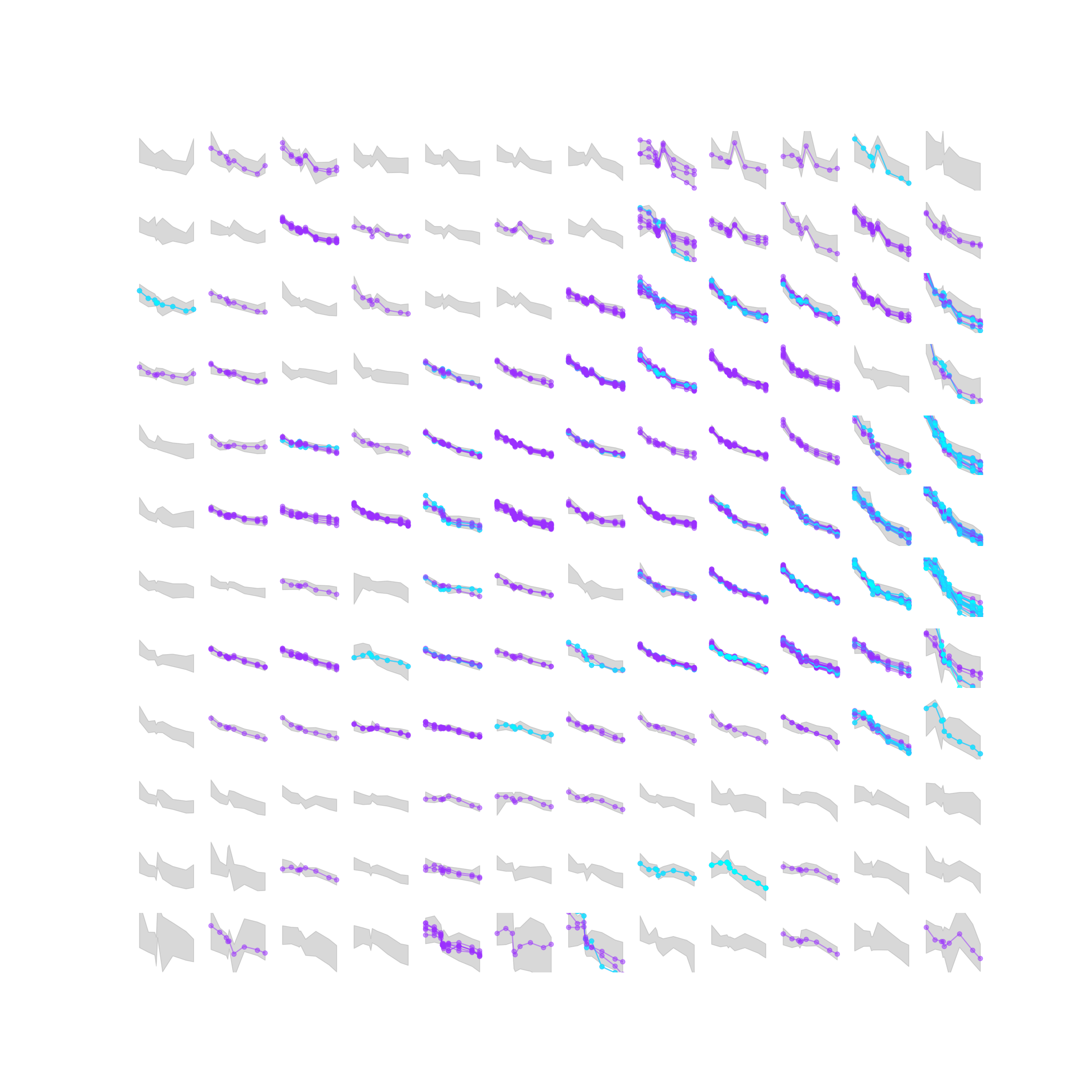}
\caption{Similar to Figure \ref{fig:SEDSOM1} on the redshift range of $2.0 < z < 2.6$.}
\label{fig:SEDSOM2}
\end{figure*}

In figure \ref{fig:massdistribution}, We map the MOSDEF targets onto the SOMs in two redshift ranges. The first range comprises 306 galaxies, including 40 identified AGNs, while the second range comprises 574 galaxies, including 123 identified AGNs. This analysis reveals that the MOSDEF targets are not randomly distributed across the map but rather tend to cluster, and that the parameter space (or the SED shapes of the galaxy population at each redshift) is not fully covered. Consequently, any conclusions drawn from these targets may not be fully representative of the entire galaxy population within these redshift ranges. This is expected, as the MOSDEF target selection is inherently mass-biased by design. Our analysis shows that more massive galaxies do not capture the full diversity of SED shapes observed in the CANDELS sample, as they tend to exhibit specific SED characteristics that differ from the broader galaxy population (figure \ref{fig:SEDSOM1} and \ref{fig:SEDSOM2}). A few important considerations should be noted. First, whether the SOM is fully populated by the mapped sample depends on the size of the grid. While reducing the size of the SOM grid can lead to a more filled map, it also causes galaxies mapped to neighboring pixels to be grouped into fewer cells, potentially leaving some SED shapes unrepresented. This effect is illustrated in Figure \ref{fig:SEDSOM1} and \ref{fig:SEDSOM2}, where a map with five times fewer cells is shown.  This detailed mapping allowed for an examination of how SED shapes vary depending on their location within the SOM. As shown in these figures, SOM reveals variability in the photometric errors across different cells; some cells exhibit higher errors while others show lower errors. This variability in error rates impacts the SOM’s ability to accurately map galaxies. When a galaxy is not mapped effectively due to high photometric errors or poor spectroscopic quality, it tends to be positioned towards the boundaries or outer cells of the map. Consequently, the SED shapes near the edges of the SOM are generally thicker and less precise compared to those at the center. Additionally, it is important to recognize that while the SOM effectively learns the manifold and covers the data space, it does not necessarily preserve the original density of observations used during training. If the density of samples is a critical factor for scientific analysis, an alternative algorithm that is sensitive to density should be considered.

As can be seen in Figure \ref{fig:massdistribution}, although our SOMs are trained on SED shapes and colors, there is a spatial trend with median stellar mass, and the MOSDEF targets tend to cluster in cells corresponding to higher median stellar masses. This observation aligns with previous studies, such as \citet{Runco2022}, which compared MOSDEF targets with a sample of star-forming galaxies at the same redshift observed in the KBSS-MOSFIRE survey \citep{strom2017}. These studies similarly identified differences between the two samples, attributed to their distinct selection functions, with MOSDEF galaxies encompassing more massive systems with redder rest-frame UV colors compared to those in the KBSS-MOSFIRE survey. (\citealt{Shivaei2015})

Similar to the application of SOMs in weak-lensing redshift calibration efforts (e.g., \citealt{Masters2019}), where non-filled regions of the SOM—corresponding to areas without prior spectroscopic observations across all redshifts—are targeted for follow-up spectroscopic studies, galaxy evolution research at specific redshifts can similarly integrate and map existing spectroscopic data. We note that the empty regions of the SOMs do not necessarily indicate poor spectroscopic target selection. Instead, these gaps may represent galaxy types that are more challenging to observe with MOSFIRE due to faint or absent emission lines. These reduced-dimensional spaces not only aid in sample comparison and completeness assessment but also serve as critical tools for planning follow-up observations. By identifying and addressing these gaps, we can work towards a more comprehensive understanding of the underlying physical processes at play. Additionally, the use of SOMs to learn the manifold and utilize the lower-dimensional projection aligns with many state-of-the-art deep learning frameworks, where data is passed through bottlenecks or lower-dimensional latent spaces (e.g., \citealt{Parker2024}). In both cases, whether through SOMs or deep neural networks, the goal is to capture the essential structure of the data in a compressed form that retains its critical features.

\subsection{Selections of AGN hosts}

\begin{figure*}[htbp]
\centering
  \includegraphics[trim=0cm 0cm 0cm 0cm, clip,width=0.99 \textwidth] {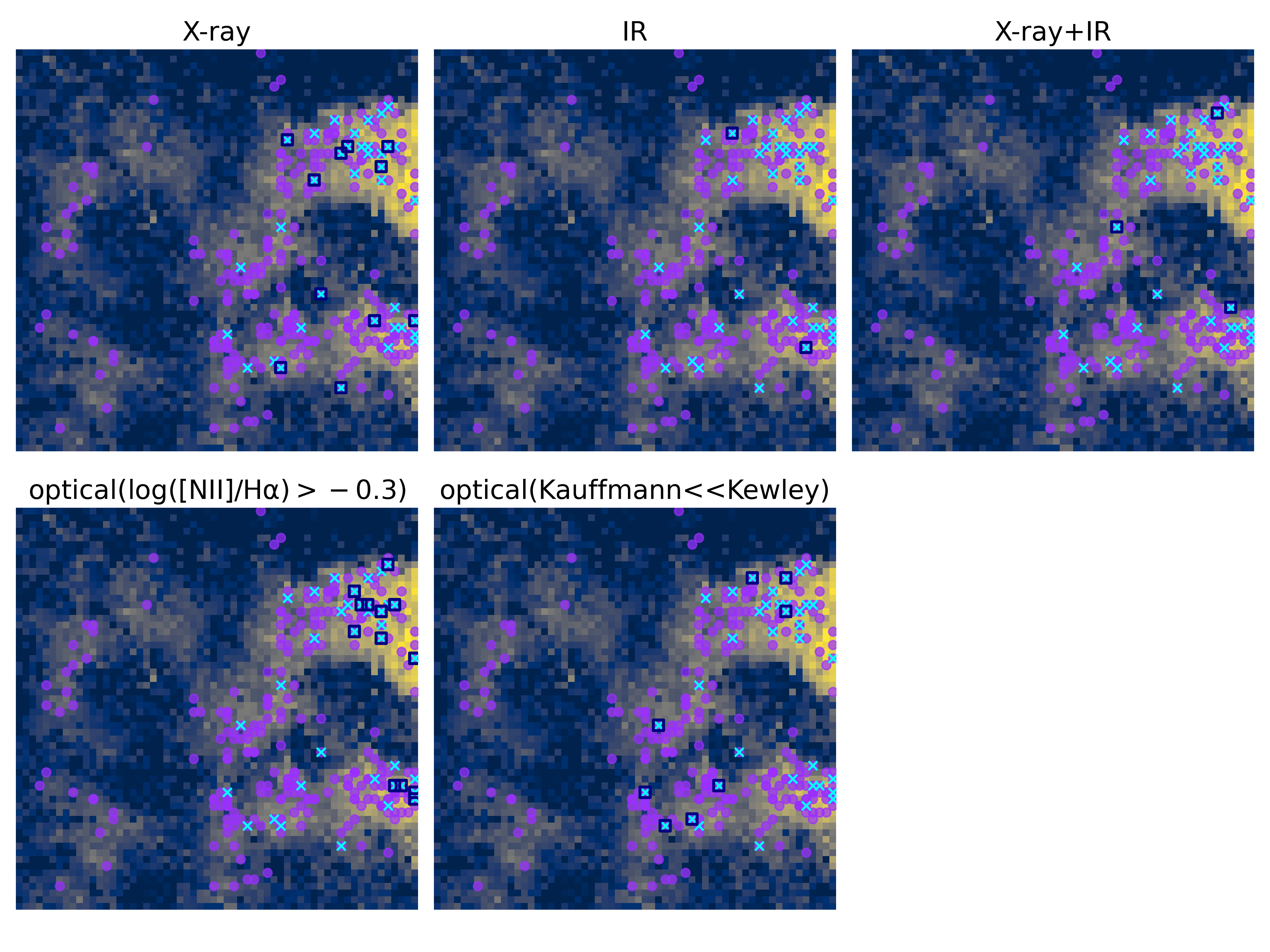}
\caption{Distribution of MOSDEF AGNs mapped on SOM in the first redshift bin at ($1.3 < z < 1.7$), shown as cyan 'x'. Blue squares around the cyan 'x' mark represent MOSDEF AGNs in each respective group. The optical (Kauffmann \( \ll \) Kewley) AGN group refers to the AGNs that lie between the \citet{Kauffmann2003} and \citet{Kewley2001} lines in the BPT \citep{Baldwin1981} diagram.
 Normal star-forming MOSDEF galaxies are indicated by purple circles.}
\label{fig:AGNz1}
\end{figure*}

\begin{figure*}[htbp]
\centering
  \includegraphics[trim=0cm 0cm 0cm 0cm, clip,width=0.99 \textwidth] {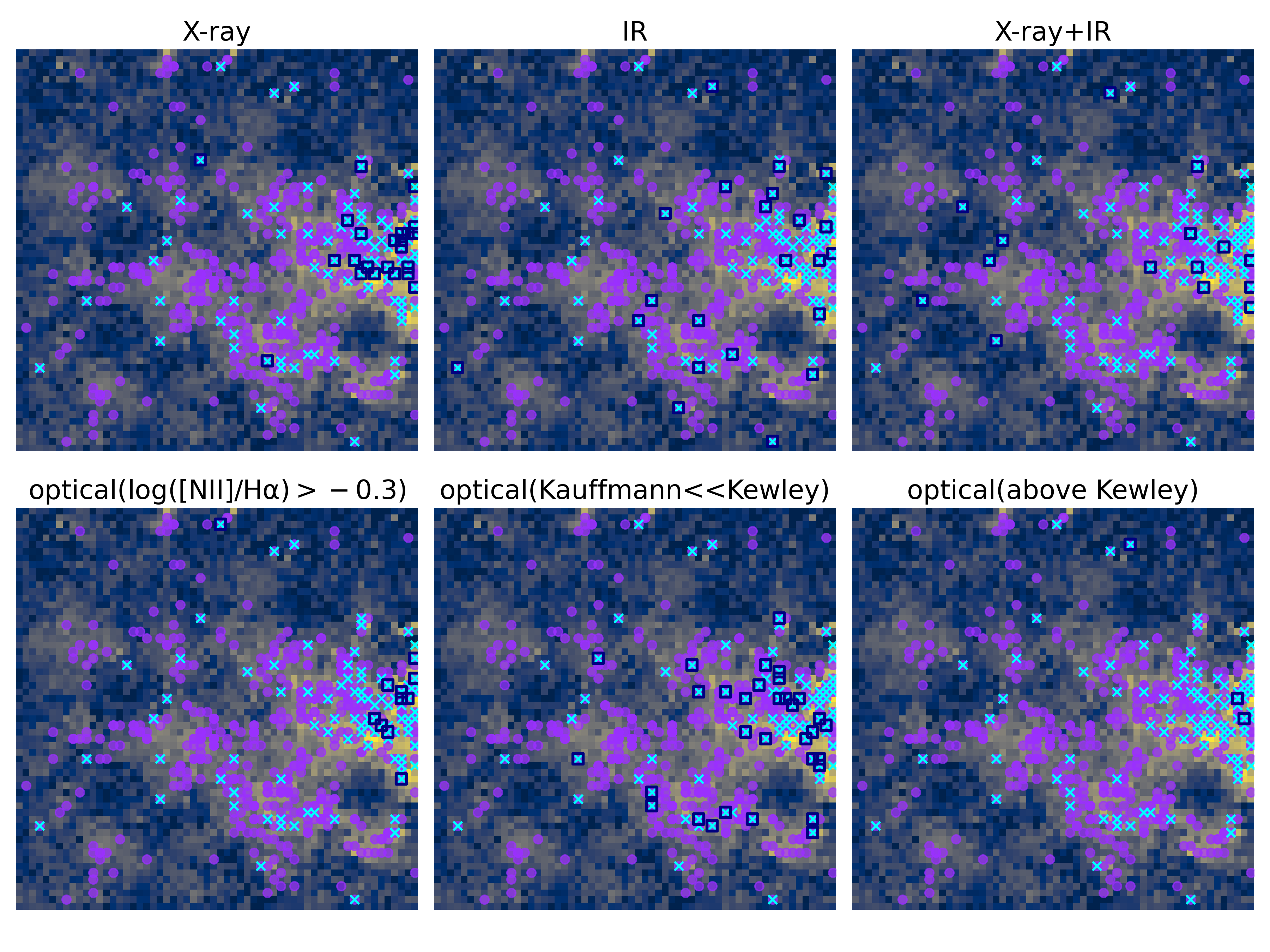}
\caption{Distribution of MOSDEF AGNs mapped on SOM in the second redshift bin at ($2.0 < z < 2.6$), shown as cyan 'x'. Blue squares around the cyan 'x' mark represent MOSDEF AGNs in each respective group. The optical (Kauffmann \( \ll \) Kewley) AGN group refers to the AGNs that lie between the \citet{Kauffmann2003} and \citet{Kewley2001} lines in the BPT \citep{Baldwin1981} diagram. The optical (above Kewley) AGN group refers to AGNs that are located above the \citet{Kewley2001} line on the BPT diagram \citep{Baldwin1981}. Normal star-forming MOSDEF galaxies are indicated by purple circles.}
\label{fig:AGNz2}
\end{figure*}

Mapping AGNs selected through different techniques onto a trained SOM is essential for understanding galaxy evolution. AGNs are closely linked to the properties of their host galaxies and their surrounding environments, making them powerful probes of the underlying physical processes at play. By determining where AGNs are positioned within the broader galaxy population and whether they cluster in specific regions of the parameter space, we can uncover selection biases and gain deeper insights into AGN activity.

Different selection methods—such as X-ray, infrared, optical, or radio observations—may favor AGNs in host galaxies with particular characteristics, such as higher stellar masses, elevated star formation rates, or larger dust content. These biases can shape our understanding of the AGN population, especially given that AGN activity may represent a transient phase in a galaxy's lifetime. Mapping AGNs onto a SOM enables a visual and quantitative assessment of these biases, helping to determine whether certain techniques preferentially identify AGNs in specific types of galaxies. Additionally, this approach allows us to explore whether AGNs are uniformly distributed across the galaxy population or tend to cluster in regions corresponding to specific stages of galaxy evolution, thereby enhancing our understanding of the relationship between AGN activity and host galaxy properties.

We mapped AGN hosts, previously identified in MOSDEF studies (\citealt{Azadi2017}), onto our SOMs within two redshift ranges (see Figure \ref{fig:massdistribution}). All of these AGNs are classified as Type 2, characterized by obscured central regions where surrounding dust and gas block direct observation of the broad-line region, resulting in emission that is predominantly influenced by the host galaxy in the UV-optical range (e.g., \citealt{Zou2019}), making them difficult to detect. The initial identification of these AGNs relied on one or more of the following methods: X-ray detections, mid-infrared (MIR) imaging, and optical emission line ratios.

The X-ray AGNs in the MOSDEF sample were identified using Chandra imaging, with exposure depths of 4 Ms in GOODS-S, 2 Ms in GOODS-N, 800 ks in EGS, and 160 ks in COSMOS. Detection in the hard X-ray band (2–10 keV) was crucial for including moderately obscured AGNs with column densities of (\( \text{N}_{\text{H}} \sim 10^{22\text{-}24} \, \text{cm}^{-2} \)). However, hard X-rays cannot penetrate Compton-thick regions (\( \text{N}_{\text{H}} > 10^{24} \, \text{cm}^{-2} \)). In such cases, nuclear emission is absorbed, reprocessed by surrounding dust, and re-emitted at MIR wavelengths, making MIR imaging a valuable tool for identifying these heavily obscured AGNs. Moreover, the MOSDEF survey utilized emission lines at 3700–7000 Å, allowing for AGN identification via optical diagnostics like the BPT  diagram (\citealt{Baldwin1981}). 

In Figures \ref{fig:massdistribution}, \ref{fig:SEDSOM1},and \ref{fig:SEDSOM2} , MOSDEF AGNs are highlighted in cyan. It is evident that in both redshift ranges, these AGNs cluster in regions of the SOM associated with higher median stellar masses, even more prominently than the overall MOSDEF targets. This pattern aligns with previous studies, which have consistently shown that AGN identification across all wavelengths is biased against low-mass galaxies. Notably, this bias is a result of observational selection effects (e.g., \citealt{Kauffmann2003,Xue2010,Aird2012}).

Figures \ref{fig:AGNz1} and \ref{fig:AGNz2}, show different groups of AGN selections mapped onto the SOMs for the first and second redshift ranges, respectively. Beginning with the X-ray selected AGNs, the variation in the depth of Chandra observations across our fields, coupled with the changing effective depth within a single field, results in a nonuniform flux limit. Consequently, X-ray imaging may fail to detect AGNs identified at other wavelengths. Previous studies have shown that X-ray selection can identify AGNs with low specific accretion rates, leading to a selection bias towards massive host galaxies (\citealt{Azadi2017}). Significantly, the SOM maps X-ray AGNs into the more massive cells, confirming the results of earlier studies.

\begin{figure*}[!htbp]
\centering
  \includegraphics[trim=0cm 0cm 0cm 0cm, clip,width=0.49\textwidth] {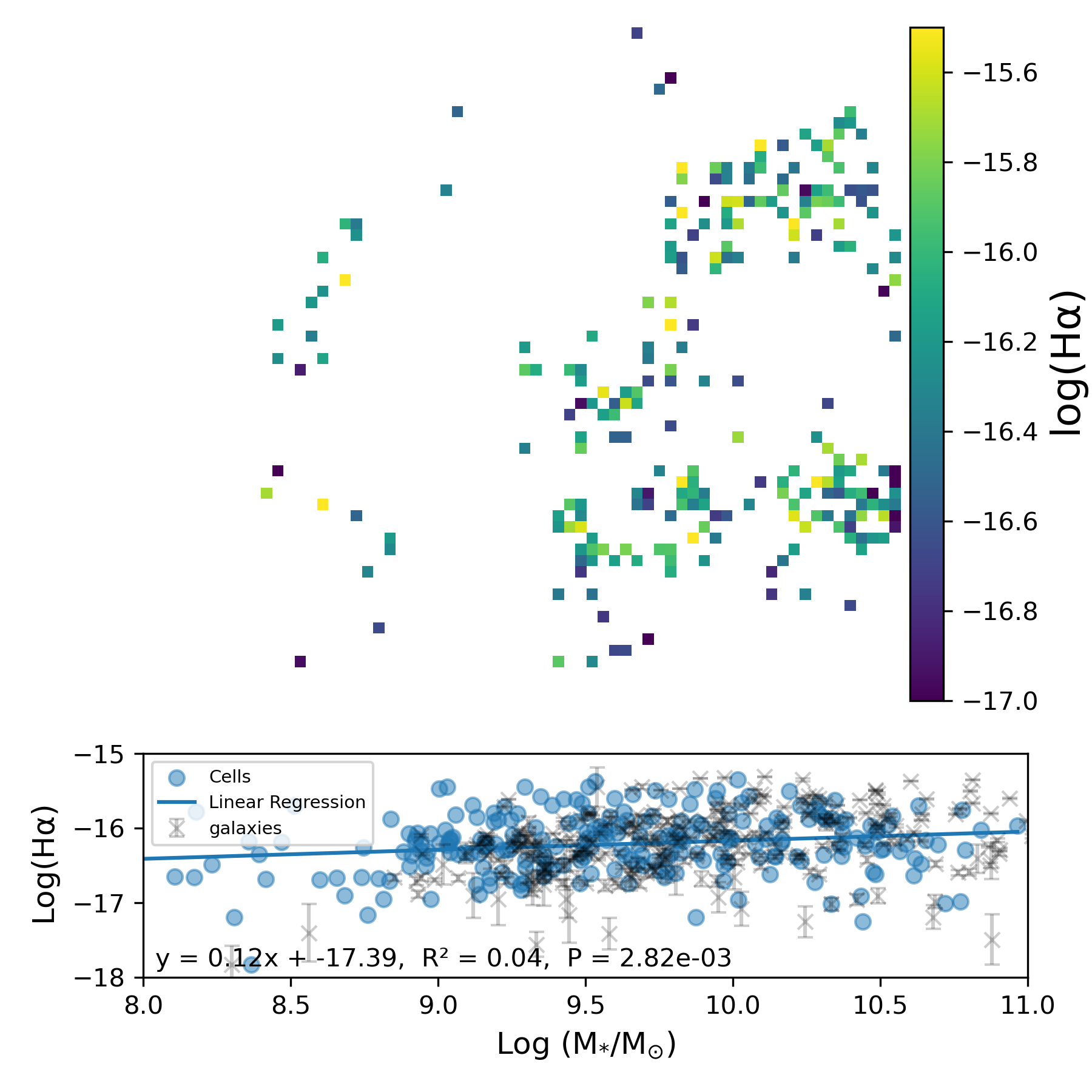}
  \includegraphics[trim=0cm 0cm 0cm 0cm, clip,width=0.49\textwidth] {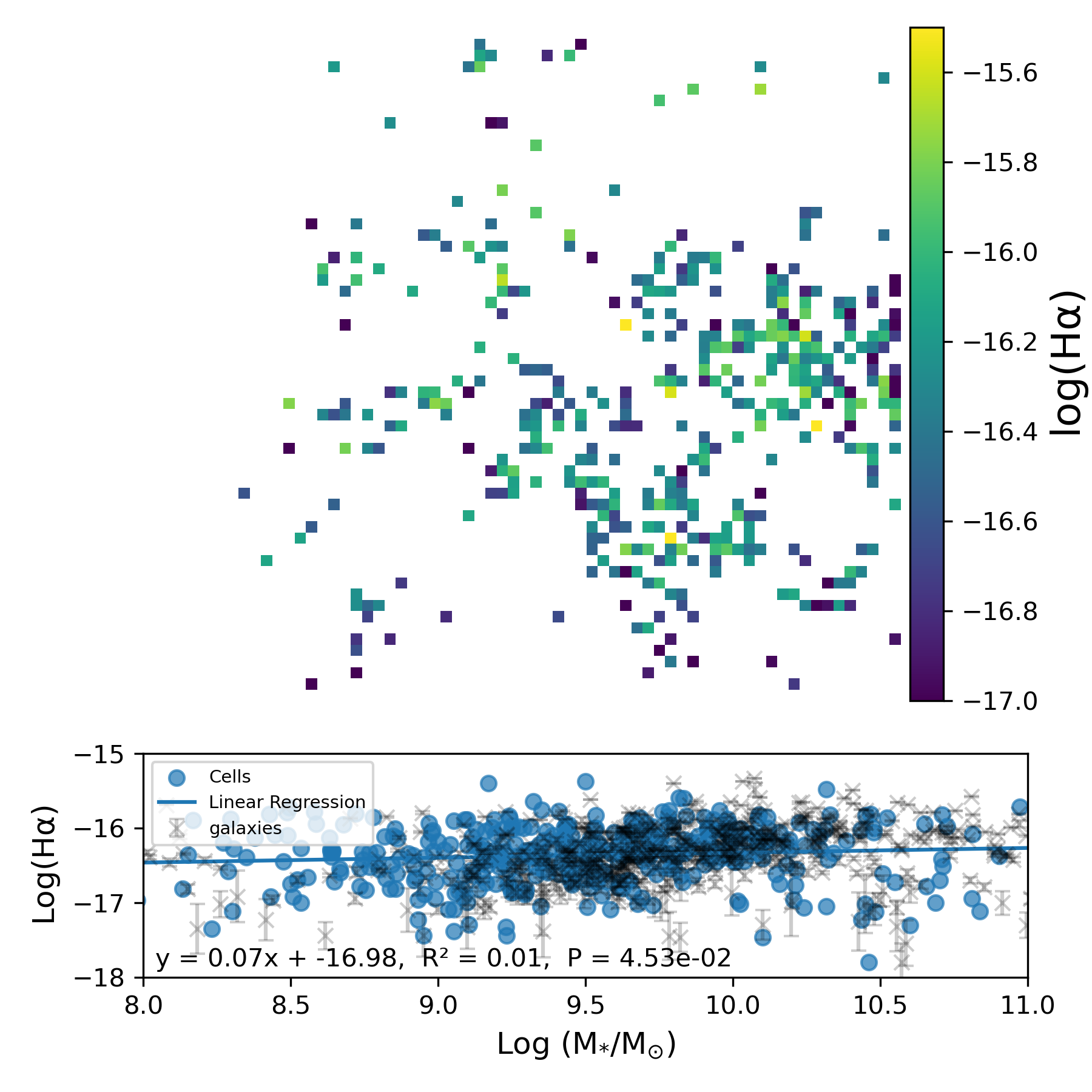}
\caption{H$\alpha$ flux mapped to the SOM cells for two redshift bins: \textbf{Left}: $1.3 < z < 1.7$, \textbf{Right}: $2.0 < z < 2.6$. Each blue point in the scatter plots represents
a SOM cell. Stellar mass values are derived from the CANDELS training sample, while H$\alpha$ flux is obtained from MOSDEF spectroscopic sample.}
\label{fig:Halpha}
\end{figure*}

\begin{figure*}[htbp]
\centering
\includegraphics[trim=0cm 0cm 0cm 0cm, clip,width=0.99 \textwidth] {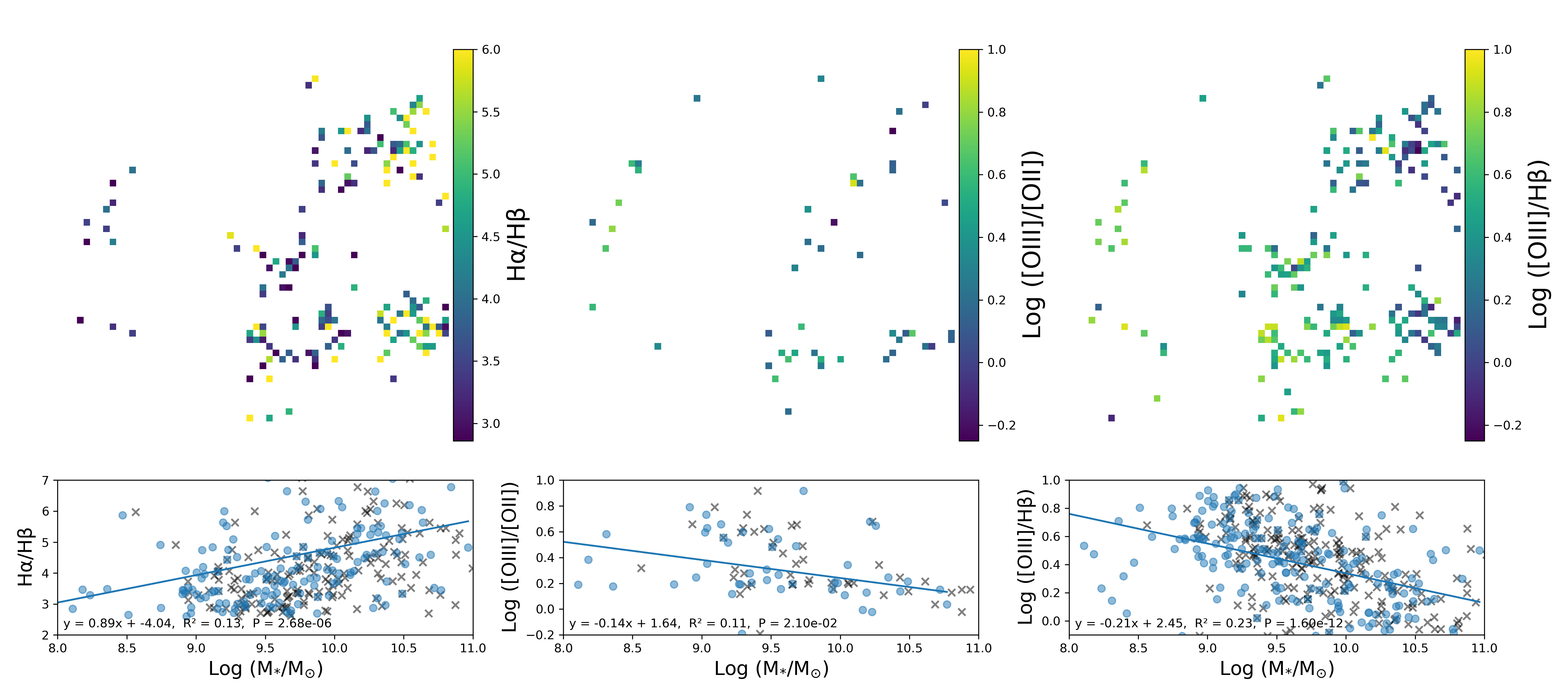}
  \includegraphics[trim=0cm 0cm 0cm 0cm, clip,width=0.99 \textwidth] {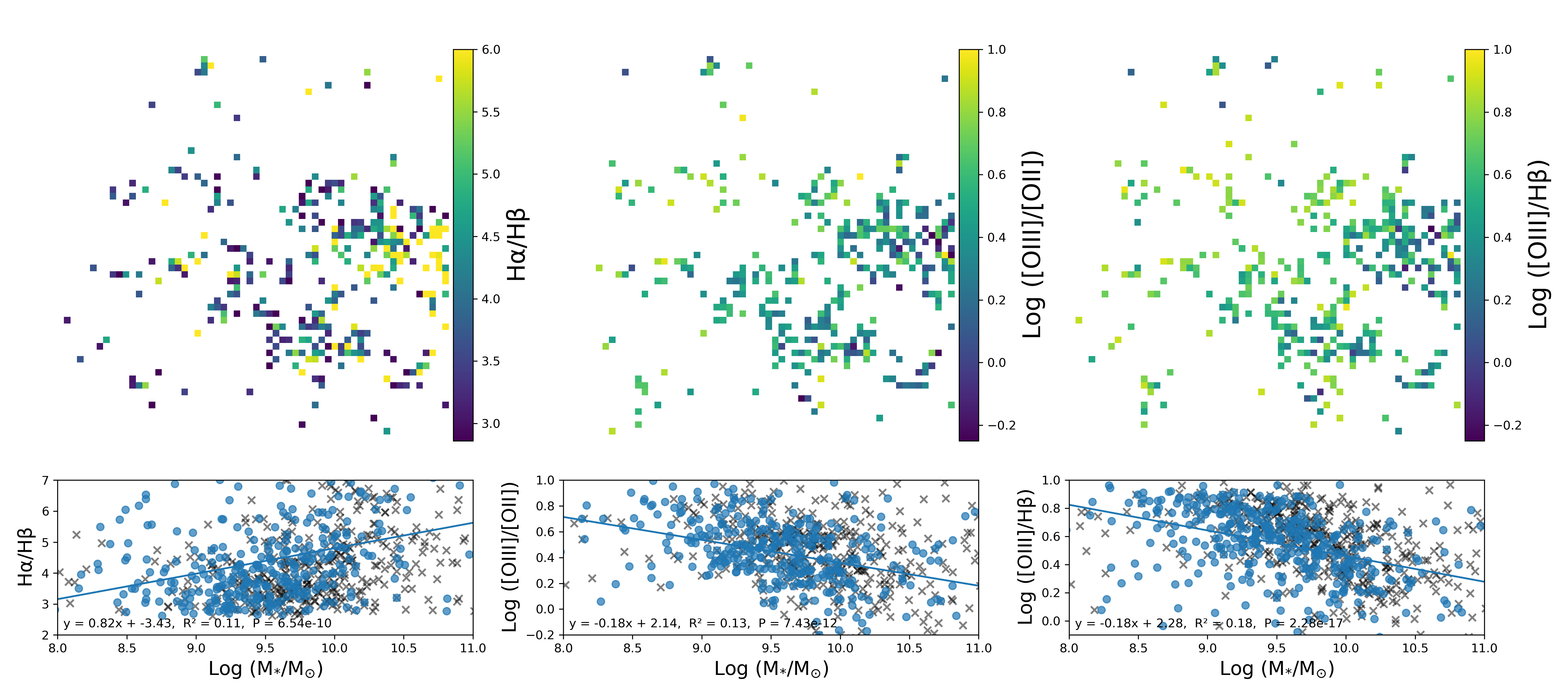}
\caption{Nebular emission line ratios of the MOSDEF galaxies in the $1.3 < z < 1.7$ (\textbf{top row}) and $2.0 < z < 2.6$ (\textbf{bottom row}) redshift ranges. From left to right, the H$\alpha$/H$\beta$, [O{\sc iii}]/[O{\sc ii}], and [O{\sc iii}]/H$\beta$ are plotted. The scatter plots show the relation with median stellar mass of each cell (blue circles) and individual mosdef galaxies (black crosses).}
\label{fig:lineratios}
\end{figure*}

\begin{figure*}[htbp]
\centering
\begin{minipage}[b]{0.49\textwidth}
    \centering
    \includegraphics[trim=0cm 0cm 0cm 0cm, clip, width=\textwidth]{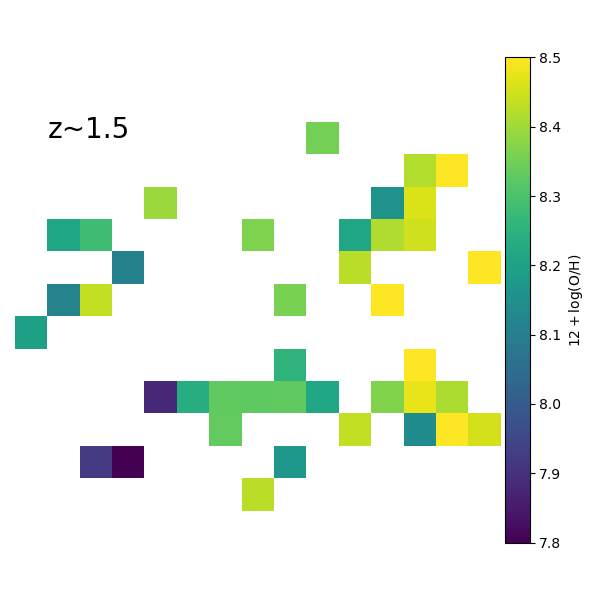}
    \parbox{0.95\textwidth}{\centering (a) $1.3 < z < 1.7$}
\end{minipage}
\hfill
\begin{minipage}[b]{0.49\textwidth}
    \centering
    \includegraphics[trim=0cm 0cm 0cm 0cm, clip, width=\textwidth]{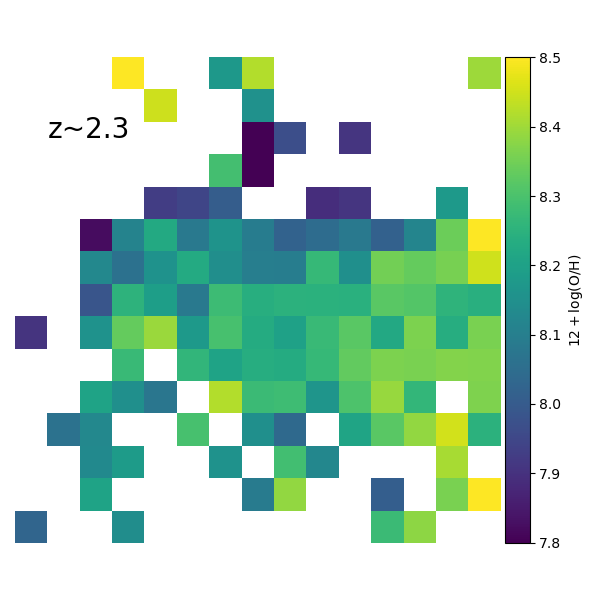}
    \parbox{0.95\textwidth}{\centering (b) $2.0 < z < 2.6$}
\end{minipage}
\caption{The gas phase metallicity distribution on a 15x15 SOM.}
\label{fig:mzr}
\end{figure*}

To identify IR AGNs, MOSDEF has used the \cite{Donley2012} selection criteria which sets a detection limit and color criteria on the four Spitzer/IRAC bands. This selection is more strict compared to the more commonly used \cite{Stern2005}, and hence finds fewer AGNs with less contamination from star-forming targets. \cite{Azadi2017} has shown that within the MOSDEF survey, the IR selected AGN are commonly found in host galaxies that are less dusty, have younger stellar populations, and exhibit higher star formation rates (SFRs). This trend is due to the stellar mass selection biases for IR AGN, as these AGN are frequently identified in galaxies with lower stellar mass. Compared to X-ray or optical AGN, these lower mass galaxies typically have less dust and younger stellar populations than higher mass galaxies. The mentioned pattern can be seen more clearly in our second redshift range SOM, where a higher number of AGNs are observed.

Optical AGN identification tends to select lower accretion rate AGNs that might not be detected at other wavelengths, primarily due to the high stellar masses of their host galaxies. Similar to X-ray AGNs, optical AGNs are often found in dusty galaxies with older stellar populations and moderate star formation rates (e.g., \citealt{Coil2015}). This selection bias towards higher mass galaxies results in higher dust content and higher metallicities, making it difficult to identify low mass, low metallicity host galaxies (\citealt{Groves2006}). Despite the small MOSDEF sample, this bias against lower mass galaxies with higher SFRs is consistent with other findings from optical AGN studies at lower redshifts (e.g., \citealt{Kauffmann2003,Trump2015}). While there were no targets sitting in the fully AGN region of the BPT diagram (i.e., above the Kewley line \citealt{Kewley2004}) in the first redshift range and only two cells with such targets in the $2.0<z<2.6$. In the composite region, the mentioned relation between different optical AGNs and their host galaxy mass is observed. 

Given the clustering of AGNs in our SOM analysis, this approach shows significant promise for selecting AGNs of different types based only on rest-frame UV-optical SEDs, particularly for future large or all-sky surveys. However, since this sample was small, tests of completeness and accuracy are not particularly meaningful at this stage. Future unbiased spectroscopic follow-ups will be crucial to fully assess and refine the method's effectiveness in capturing the diverse AGN population across various host galaxy properties.

\subsection{Strong Optical Emission Lines}

Scaling relations are fundamental tools in understanding galaxy evolution, providing key insights into how various physical properties of galaxies are interconnected. Relations like the star formation main sequence (e.g., \citealt{Noeske2007,Shivaei2015,Speakle2014}) or the mass-metallicity relation (MZR; e.g., \citealt{Tremonti2004,Finlator2008,Curti2024}) show consistent patterns across large samples of galaxies, offering a framework to study how galaxies grow, form stars, and enrich their interstellar medium over cosmic time. These relations are crucial for interpreting the complex interplay of gas accretion, star formation, feedback processes, and galaxy mergers that drive galaxy evolution.

Nebular emission lines and their ratios, derived from spectroscopy, offer powerful diagnostics that significantly enhance scaling relations by probing key physical properties of galaxies. Emission lines, such as H$\alpha$ originating from ionized gas in H{\sc ii} regions, provide direct measurements of star formation rate. The Balmer decrement (H$\alpha$/H$\beta$) is particularly important as it measures nebular dust extinction, allowing for more accurate corrections of observed line fluxes and reliable estimates of intrinsic galaxy properties. In combination with other lines, such as [O{\sc iii}] and [O{\sc ii}], ratios like [O{\sc iii}]/H$\beta$ and [O{\sc iii}]/[O{\sc ii}] offer precise assessments of a galaxy’s ionization state and gas-phase chemical composition. 

Spectroscopy provides valuable data but it is expensive, and it is of value to extend the learned diagnostics and relations to a larger population. Also, the scaling relations typically focus and relate only on two or three key properties. For instance, the MZR which describes the tight correlation between stellar mass and metallicity, has been further refined by adding SFR as a third parameter, the so-called fundamental metallicity relation (FMR; \citealt{Mannucci2010}). FMR shows less evolution with redshift compared to MZR, whose intercept evolves strongly with cosmic time (e.g., \citealt{Lara2010}, \citealt{Nakajima2023}). Additional studies have explored the scatter in the MZR by linking it to other galaxy properties, such as dust content (e.g., \citealt{Chartab2022}) or galaxy size (e.g., \citealt{Almeida2018}).

Here, by mapping the nebular line ratios onto the SOM, we explore the scaling relations with the full SED shape, rather than relying on one or two physical properties at a time. Moreover, discrepancies between different studies often arise due to variations in sample selection, observational techniques, or methodological approaches. By utilizing the SOM, one can visually identify clusters and trends that reveal hidden correlations between the SED shape and nebular line diagnostics among different samples. 

The nebular line fluxes in the MOSDEF survey were corrected for slit loss \cite{Kriek2015} and extinction using the Milky Way extinction curve (\citealt{Cardelli1989}) and the Balmer decrement, when available. For targets where either H$\alpha$ or H$\beta$ was unavailable, a conversion based on stellar extinction was applied (\citealt{Sanders2021}). In Figure \ref{fig:Halpha}, we map the H$\alpha$ fluxes of the MOSDEF targets onto two SOMs, revealing numerous pixels with no measurements, either due to the lack of MOSDEF sampling or the absence of strong emission lines detectable in the MOSDEF observations. While higher overall H$\alpha$ fluxes are observed in areas of the SOM corresponding to higher stellar masses, the trends in H$\alpha$ flux across the SOM are relatively weak. This is partly due to the small sample size relative to the SOM resolution and the fact that the SOM was trained on normalized SED shapes. In the lower subplots, we also show the relationship between H$\alpha$ flux and the median stellar mass in each cell, overlaid with individual MOSDEF galaxies. A linear fit to the cells shows consistent slope and intercept across two redshift ranges.

The spatial trends of line ratios on the SOM are more pronounced, as shown in Figure \ref{fig:lineratios}. From left to right, we present the Balmer decrement, the [O{\sc iii}]/[O{\sc ii}], and the [O{\sc iii}]/H$\beta$ ratios for the two redshift ranges. The expected correlations, where regions with higher galaxy mass tend to display lower [O{\sc iii}]/H$\beta$ and [O{\sc iii}]/[O{\sc ii}] ratios alongside higher H$\alpha$/H$\beta$ ratios, suggest that more massive galaxies typically have higher metal content, which influences the ionization levels of their gas. The increased presence of metals generally enhances cooling processes, thus lowering the ionization potential in these regions. Note that the scatter plots below each SOM plot are just there to be compared with expectations from the trends of the line ratio with stellar mass. And even with no information about the stellar mass one can evidently see the spatial relation of the line ratios on the SOM.

The [O{\sc iii}]/[O{\sc ii}] ratio (O32) was then converted to gas-phase metallicity using the relation from \cite{Bian2018} (i.e., $12+ \log(O/H) = 8.54 - 0.59 \times O32$), following the approach used by \citealt{Sanders2021}. It is important to note that this relation is calibrated for star-forming galaxies and may not be universally applicable. Additionally, O32 primarily serves as a tracer of the ionization parameter (e.g., \citealt{Kewley2002, Nagao2006}), with its strong dependence on metallicity being secondary and driven by the correlation between ionization parameter and metallicity (\citealt{Maiolino2019}). However, to visualize the MZR or a metallicity trend on the SOM, Figure \ref{fig:mzr} presents the distribution of gas-phase metallicity on a smaller $15\times15$ grid. This smaller grid is used to enhance the visual clarity of the trends between the line ratio and the SED shape. Despite the lower redshift bin having significantly fewer observations covering both [O{\sc iii}] and [O{\sc ii}], the observed trends are consistent across redshift.

This opens up new avenues for future studies on the MZR, FMR, and their redshift evolution, as well as their dependence on different galaxy samples. For example, examining galaxies with and without AGNs, \cite{Li2024} found that while low-mass ($\log(M_{*}/M_{\odot})<10.5$) AGN hosts exhibit metallicities approximately 0.2 dex higher than non-hosts at the same stellar masses, this MZR offset in the FMR is significantly smaller. They suggest that the larger deviation in the MZR is due to AGN-host galaxies having systematically lower SFRs at fixed stellar mass. Another potential use case is enhanced selection of Lyman continuum leakers. The most promising low-redshift analogs to reionization-era Lyman continuum-leaking galaxies are shown to be Lyman alpha emitters with low gas-phase metallicity (e.g., \citealt{Nakajima2016,Trainor2016,Fletcher2019}) and high O32 ratios (\citealt{Nakajima2020}). With nebular line ratios mapped onto the SOM, future searches for Lyman continuum leakage in galaxies may not require rest-frame optical spectra covering these oxygen lines. Instead, they can narrow the search space using photometric optical+NIR SED data.

\section{Summary}

This study illustrates the application of manifold learning and dimensionality reduction in analyzing high-dimensional astronomical datasets. By estimating missing fluxes for galaxies in the CANDELS fields, we constructed SEDs for individual galaxies using an 8-dimensional color space covering: $u-g$, $g-r$, $r-i$, $i-z$, $z-Y$, $Y-J$, $J-H$, and $H-H2$. The SEDs were used to train the SOMs in two redshift intervals corresponding to those from the MOSDEF sample: $1.3 < z < 1.7$ (9,976 galaxies) and $2.0 < z < 2.6$ (14,130 galaxies). In this SOM configuration, neighboring cells contain galaxies with similar SEDs, with objects with very close SEDs residing in the same cell. The main findings are summarized as follows:

\begin{itemize}
    \item MOSDEF galaxies are not uniformly distributed across the SOM and, as a result, do not fully represent the diversity of SEDs in the CANDELS galaxy populations. Instead, they tend to cluster in regions associated with higher stellar masses, reflecting the mass-biased design of the MOSDEF target selection. The SOM confirms this selection bias by learning the characteristic SEDs of massive galaxies, further highlighting the correlation between SED and stellar mass. Additionally, we observe that galaxies with higher photometric errors are predominantly located at the edges of the SOM, where the CANDELS SEDs show increased thickness, indicative of larger error bars.
    \item The SOMs is used to confirm known observational biases and selection effects inherent in galaxy surveys. For example, AGNs were preferentially identified to be associated with host galaxies with higher mass, lower dust and younger stellar populations. The mapping of the AGNs into more massive cells agrees with previous finding that optical and X-ray selected AGNs reside in more massive host galaxies. Selecting targets from these regions to identify AGNs does not guarantee a positive detection. However, this method substantially increases the likelihood of success compared to a random or less-informed selection process, as it incorporates prior knowledge into the target selection.
    \item We mapped the MOSDEF galaxies onto the SOM, trained with broadband photometric features (i.e., colors) from the CANDELS data. The SOM was color-coded based on the spectroscopic features from MOSDEF galaxies. To test the effectiveness of the SOMs in mapping targets, we also plotted the emission line ratios from MOSDEF galaxies diagnostic of physical parameters (e.g., H$\alpha$, [O{\sc iii}]/H$\beta$, [O{\sc iii}]/[O{\sc ii}], H$\alpha$/H$\beta$) against the corresponding stellar mass from each SOMs cell. Our analysis confirmed the well-established mass-metallicity relation, showing that more massive galaxies tend to have lower [O{\sc iii}]/H$\beta$ and [O{\sc iii}]/[O{\sc ii}] ratios, as well as higher H$\alpha$ flux and H$\alpha$/H$\beta$ ratios.
\end{itemize}

\bibliography{references.bib}

\begin{thebibliography}{}
\expandafter\ifx\csname natexlab\endcsname\relax\def\natexlab#1{#1}\fi
\providecommand{\url}[1]{\href{#1}{#1}}
\providecommand{\dodoi}[1]{doi:~\href{http://doi.org/#1}{\nolinkurl{#1}}}
\providecommand{\doeprint}[1]{\href{http://ascl.net/#1}{\nolinkurl{http://ascl.net/#1}}}
\providecommand{\doarXiv}[1]{\href{https://arxiv.org/abs/#1}{\nolinkurl{https://arxiv.org/abs/#1}}}

\bibitem[{Abareshi {et~al.}(2022)Abareshi, Aguilar, Ahlen, Alam, Alexander, Alfarsy, Allen, Prieto, Alves, Ameel, Armengaud, Asorey, Aviles, Bailey, Balaguera-Antolínez, Ballester, Baltay, Bault, Beltran, Benavides, BenZvi, Berti, Besuner, Beutler, Bianchi, Blake, Blanc, Blum, Bolton, Bose, Bramall, Brieden, Brodzeller, Brooks, Brownewell, Buckley-Geer, Cahn, Cai, Canning, Capasso, Rosell, Carton, Casas, Castander, Cervantes-Cota, Chabanier, Chaussidon, Chuang, Circosta, Cole, Cooper, Costa, Cousinou, Cuceu, Davis, Dawson, Cruz-Noriega, Macorra, Mattia, Costa, Demmer, Derwent, Dey, Dey, Dhungana, Ding, Dobson, Doel, Donald-McCann, Donaldson, Douglass, Duan, Dunlop, Edelstein, Eftekharzadeh, Eisenstein, Enriquez-Vargas, Escoffier, Evatt, Fagrelius, Fan, Fanning, Fawcett, Ferraro, Ereza, Flaugher, Font-Ribera, Forero-Romero, Frenk, Fromenteau, Gänsicke, Garcia-Quintero, Garrison, Gaztañaga, Gerardi, Gil-Marín, A~Gontcho, Gonzalez-Morales, Gonzalez-de Rivera, Gonzalez-Perez, Gordon, Graur, Green, Grove,
  Gruen, Gutierrez, Guy, Hahn, Harris, Herrera, Herrera-Alcantar, Honscheid, Howlett, Huterer, Iršič, Ishak, Jelinsky, Jiang, Jimenez, Jing, Joyce, Jullo, Juneau, Karaçaylı, Karamanis, Karcher, Karim, Kehoe, Kent, Kirkby, Kisner, Kitaura, Koposov, Kovács, Kremin, Krolewski, L’Huillier, Lahav, Lambert, Lamman, Lan, Landriau, Lane, Lang, Lange, Lasker, Guillou, Leauthaud, Van~Suu, Levi, Li, Magneville, Manera, Manser, Marshall, Martini, McCollam, McDonald, Meisner, Mena-Fernández, Meneses-Rizo, Mezcua, Miller, Miquel, Montero-Camacho, Moon, Moustakas, Mueller, Muñoz-Gutiérrez, Myers, Nadathur, Najita, Napolitano, Neilsen, Newman, Nie, Ning, Niz, Norberg, Noriega, O’Brien, Obuljen, Palanque-Delabrouille, Palmese, Zhiwei, Pappalardo, PENG, Percival, Perruchot, Pogge, Poppett, Porredon, Prada, Prochaska, Pucha, Pérez-Fernández, Pérez-Ràfols, Rabinowitz, Raichoor, Ramirez-Solano, Ramírez-Pérez, Ravoux, Reil, Rezaie, Rocher, Rockosi, Roe, Roodman, Ross, Rossi, Ruggeri, Ruhlmann-Kleider, Sabiu,
  Safonova, Said, Saintonge, Catonga, Samushia, Sanchez, Saulder, Schaan, Schlafly, Schlegel, Schmoll, Scholte, Schubnell, Secroun, Seo, Serrano, Sharples, Sholl, Silber, Silva, Sirk, Siudek, Smith, Sprayberry, Staten, Stupak, Tan, Tarlé, Tie, Tojeiro, Ureña-López, Valdes, Valenzuela, Valluri, Vargas-Magaña, Verde, Walther, Wang, Wang, Weaver, Weaverdyck, Wechsler, Wilson, Yang, Yu, Yuan, Yèche, Zhang, Zhang, Zhao, Zhou, Zhou, Zou, Zou, Zou, \& Zu}]{Abareshi2022}
Abareshi, B., Aguilar, J., Ahlen, S., {et~al.} 2022, The Astronomical Journal, 164, 207, \dodoi{10.3847/1538-3881/ac882b}

\bibitem[{{Ackermann} {et~al.}(2018){Ackermann}, {Schawinski}, {Zhang}, {Weigel}, \& {Turp}}]{merger2018}
{Ackermann}, S., {Schawinski}, K., {Zhang}, C., {Weigel}, A.~K., \& {Turp}, M.~D. 2018, \mnras, 479, 415, \dodoi{10.1093/mnras/sty1398}

\bibitem[{Acquaviva(2015)}]{Acquaviva2015}
Acquaviva, V. 2015, Monthly Notices of the Royal Astronomical Society, 456, 1618, \dodoi{10.1093/mnras/stv2703}

\bibitem[{{Aird} {et~al.}(2012){Aird}, {Coil}, {Moustakas}, {Blanton}, {Burles}, {Cool}, {Eisenstein}, {Smith}, {Wong}, \& {Zhu}}]{Aird2012}
{Aird}, J., {Coil}, A.~L., {Moustakas}, J., {et~al.} 2012, \apj, 746, 90, \dodoi{10.1088/0004-637X/746/1/90}

\bibitem[{Azadi {et~al.}(2017)Azadi, Coil, Aird, Reddy, Shapley, Freeman, Kriek, Leung, Mobasher, Price, Sanders, Shivaei, \& Siana}]{Azadi2017}
Azadi, M., Coil, A.~L., Aird, J., {et~al.} 2017, The Astrophysical Journal, 835, 27, \dodoi{10.3847/1538-4357/835/1/27}

\bibitem[{Baldwin {et~al.}(1981)Baldwin, Phillips, \& Terlevich}]{Baldwin1981}
Baldwin, A., Phillips, M.~M., \& Terlevich, R. 1981, Publications of the Astronomical Society of the Pacific, 93, 817, \dodoi{10.1086/130930}

\bibitem[{BALL \& BRUNNER(2010)}]{Ball2010}
BALL, N.~M., \& BRUNNER, R.~J. 2010, International Journal of Modern Physics D, 19, 1049, \dodoi{10.1142/S0218271810017160}

\bibitem[{Baron(2019)}]{baron2019}
Baron, D. 2019, Machine Learning in Astronomy: a practical overview.
\newblock \doarXiv{1904.07248}

\bibitem[{Barro {et~al.}(2019)Barro, Pérez-González, Cava, Brammer, Pandya, Moral, Esquej, Domínguez-Sánchez, Pampliega, Guo, Koekemoer, Trump, Ashby, Cardiel, Castellano, Conselice, Dickinson, Dolch, Donley, Briones, Faber, Fazio, Ferguson, Finkelstein, Fontana, Galametz, Gardner, Gawiser, Giavalisco, Grazian, Grogin, Hathi, Hemmati, Hernán-Caballero, Kocevski, Koo, Kodra, Lee, Lin, Lucas, Mobasher, McGrath, Nandra, Nayyeri, Newman, Pforr, Peth, Rafelski, Rodríguez-Munoz, Salvato, Stefanon, van~der Wel, Willner, Wiklind, \& Wuyts}]{Barro2019}
Barro, G., Pérez-González, P.~G., Cava, A., {et~al.} 2019, The Astrophysical Journal Supplement Series, 243, 22, \dodoi{10.3847/1538-4365/ab23f2}

\bibitem[{{Bian} {et~al.}(2018){Bian}, {Kewley}, \& {Dopita}}]{Bian2018}
{Bian}, F., {Kewley}, L.~J., \& {Dopita}, M.~A. 2018, \apj, 859, 175, \dodoi{10.3847/1538-4357/aabd74}

\bibitem[{{Cardelli} {et~al.}(1989){Cardelli}, {Clayton}, \& {Mathis}}]{Cardelli1989}
{Cardelli}, J.~A., {Clayton}, G.~C., \& {Mathis}, J.~S. 1989, \apj, 345, 245, \dodoi{10.1086/167900}

\bibitem[{{Carrasco Kind} \& {Brunner}(2013)}]{Kind2013}
{Carrasco Kind}, M., \& {Brunner}, R.~J. 2013, \mnras, 432, 1483, \dodoi{10.1093/mnras/stt574}

\bibitem[{{Chartab} {et~al.}(2022){Chartab}, {Cooray}, {Ma}, {Nayyeri}, {Zilliot}, {Lopez}, {Fadda}, {Herrera-Camus}, {Malkan}, {Rigopoulou}, {Sheth}, \& {Wardlow}}]{Chartab2022}
{Chartab}, N., {Cooray}, A., {Ma}, J., {et~al.} 2022, Nature Astronomy, 6, 844, \dodoi{10.1038/s41550-022-01679-y}

\bibitem[{Chartab {et~al.}(2023)Chartab, Mobasher, Cooray, Hemmati, Sattari, Ferguson, Sanders, Weaver, Stern, McCracken, Masters, Toft, Capak, Davidzon, Dickinson, Rhodes, Moneti, Ilbert, Zalesky, McPartland, Szapudi, Koekemoer, Teplitz, \& Giavalisco}]{Chartab2023}
Chartab, N., Mobasher, B., Cooray, A.~R., {et~al.} 2023, The Astrophysical Journal, 942, 91, \dodoi{10.3847/1538-4357/acacf5}

\bibitem[{Coil {et~al.}(2015)Coil, Aird, Reddy, Shapley, Kriek, Siana, Mobasher, Freeman, Price, \& Shivaei}]{Coil2015}
Coil, A.~L., Aird, J., Reddy, N., {et~al.} 2015, The Astrophysical Journal, 801, 35, \dodoi{10.1088/0004-637X/801/1/35}

\bibitem[{Collaboration {et~al.}(2009)Collaboration, Abell, Allison, Anderson, Andrew, Angel, Armus, Arnett, Asztalos, Axelrod, Bailey, Ballantyne, Bankert, Barkhouse, Barr, Barrientos, Barth, Bartlett, Becker, Becla, Beers, Bernstein, Biswas, Blanton, Bloom, Bochanski, Boeshaar, Borne, Bradac, Brandt, Bridge, Brown, Brunner, Bullock, Burgasser, Burge, Burke, Cargile, Chandrasekharan, Chartas, Chesley, Chu, Cinabro, Claire, Claver, Clowe, Connolly, Cook, Cooke, Cooray, Covey, Culliton, de~Jong, de~Vries, Debattista, Delgado, Dell'Antonio, Dhital, Stefano, Dickinson, Dilday, Djorgovski, Dobler, Donalek, Dubois-Felsmann, Durech, Eliasdottir, Eracleous, Eyer, Falco, Fan, Fassnacht, Ferguson, Fernandez, Fields, Finkbeiner, Figueroa, Fox, Francke, Frank, Frieman, Fromenteau, Furqan, Galaz, Gal-Yam, Garnavich, Gawiser, Geary, Gee, Gibson, Gilmore, Grace, Green, Gressler, Grillmair, Habib, Haggerty, Hamuy, Harris, Hawley, Heavens, Hebb, Henry, Hileman, Hilton, Hoadley, Holberg, Holman, Howell, Infante, Ivezic,
  Jacoby, Jain, R, Jedicke, Jee, Jernigan, Jha, Johnston, Jones, Juric, Kaasalainen, Styliani, Kafka, Kahn, Kaib, Kalirai, Kantor, Kasliwal, Keeton, Kessler, Knezevic, Kowalski, Krabbendam, Krughoff, Kulkarni, Kuhlman, Lacy, Lepine, Liang, Lien, Lira, Long, Lorenz, Lotz, Lupton, Lutz, Macri, Mahabal, Mandelbaum, Marshall, May, McGehee, Meadows, Meert, Milani, Miller, Miller, Mills, Minniti, Monet, Mukadam, Nakar, Neill, Newman, Nikolaev, Nordby, O'Connor, Oguri, Oliver, Olivier, Olsen, Olsen, Olszewski, Oluseyi, Padilla, Parker, Pepper, Peterson, Petry, Pinto, Pizagno, Popescu, Prsa, Radcka, Raddick, Rasmussen, Rau, Rho, Rhoads, Richards, Ridgway, Robertson, Roskar, Saha, Sarajedini, Scannapieco, Schalk, Schindler, Schmidt, Schmidt, Schneider, Schumacher, Scranton, Sebag, Seppala, Shemmer, Simon, Sivertz, Smith, Smith, Smith, Spitz, Stanford, Stassun, Strader, Strauss, Stubbs, Sweeney, Szalay, Szkody, Takada, Thorman, Trilling, Trimble, Tyson, Berg, Berk, VanderPlas, Verde, Vrsnak, Walkowicz, Wandelt, Wang,
  Wang, Warner, Wechsler, West, Wiecha, Williams, Willman, Wittman, Wolff, Wood-Vasey, Wozniak, Young, Zentner, \& Zhan}]{LSST2009}
Collaboration, L.~S., Abell, P.~A., Allison, J., {et~al.} 2009, LSST Science Book, Version 2.0.
\newblock \doarXiv{0912.0201}

\bibitem[{{Collister} \& {Lahav}(2004)}]{Collister2004}
{Collister}, A.~A., \& {Lahav}, O. 2004, \pasp, 116, 345, \dodoi{10.1086/383254}

\bibitem[{{Curti} {et~al.}(2024){Curti}, {Maiolino}, {Curtis-Lake}, {Chevallard}, {Carniani}, {D'Eugenio}, {Looser}, {Scholtz}, {Charlot}, {Cameron}, {{\"U}bler}, {Witstok}, {Boyett}, {Laseter}, {Sandles}, {Arribas}, {Bunker}, {Giardino}, {Maseda}, {Rawle}, {Rodr{\'\i}guez Del Pino}, {Smit}, {Willott}, {Eisenstein}, {Hausen}, {Johnson}, {Rieke}, {Robertson}, {Tacchella}, {Williams}, {Willmer}, {Baker}, {Bhatawdekar}, {Egami}, {Helton}, {Ji}, {Kumari}, {Perna}, {Shivaei}, \& {Sun}}]{Curti2024}
{Curti}, M., {Maiolino}, R., {Curtis-Lake}, E., {et~al.} 2024, \aap, 684, A75, \dodoi{10.1051/0004-6361/202346698}

\bibitem[{Daddi {et~al.}(2007)Daddi, Dickinson, Morrison, Chary, Cimatti, Elbaz, Frayer, Renzini, Pope, Alexander, Bauer, Giavalisco, Huynh, Kurk, \& Mignoli}]{Daddi2007}
Daddi, E., Dickinson, M., Morrison, G., {et~al.} 2007, The Astrophysical Journal, 670, 156, \dodoi{10.1086/521818}

\bibitem[{{Davidzon} {et~al.}(2019){Davidzon}, {Laigle}, {Capak}, {Ilbert}, {Masters}, {Hemmati}, {Apostolakos}, {Coupon}, {de la Torre}, {Devriendt}, {Dubois}, {Kashino}, {Paltani}, \& {Pichon}}]{Davidzon2019}
{Davidzon}, I., {Laigle}, C., {Capak}, P.~L., {et~al.} 2019, \mnras, 489, 4817, \dodoi{10.1093/mnras/stz2486}

\bibitem[{{Davidzon} {et~al.}(2022){Davidzon}, {Jegatheesan}, {Ilbert}, {de la Torre}, {Leslie}, {Laigle}, {Hemmati}, {Masters}, {Blanquez-Sese}, {Kauffmann}, {Magdis}, {Ma{\l}ek}, {McCracken}, {Mobasher}, {Moneti}, {Sanders}, {Shuntov}, {Toft}, \& {Weaver}}]{Davidzon2022}
{Davidzon}, I., {Jegatheesan}, K., {Ilbert}, O., {et~al.} 2022, \aap, 665, A34, \dodoi{10.1051/0004-6361/202243249}

\bibitem[{{Domber} {et~al.}(2022){Domber}, {Gygax}, {Aumiller}, {Whipple}, {Walker}, \& {Delker}}]{Domber2022}
{Domber}, J.~L., {Gygax}, J.~D., {Aumiller}, P., {et~al.} 2022, in Society of Photo-Optical Instrumentation Engineers (SPIE) Conference Series, Vol. 12180, Space Telescopes and Instrumentation 2022: Optical, Infrared, and Millimeter Wave, ed. L.~E. {Coyle}, S.~{Matsuura}, \& M.~D. {Perrin}, 121801O, \dodoi{10.1117/12.2633897}

\bibitem[{{Donley} {et~al.}(2012){Donley}, {Koekemoer}, {Brusa}, {Capak}, {Cardamone}, {Civano}, {Ilbert}, {Impey}, {Kartaltepe}, {Miyaji}, {Salvato}, {Sanders}, {Trump}, \& {Zamorani}}]{Donley2012}
{Donley}, J.~L., {Koekemoer}, A.~M., {Brusa}, M., {et~al.} 2012, \apj, 748, 142, \dodoi{10.1088/0004-637X/748/2/142}

\bibitem[{Doré {et~al.}(2015)Doré, Bock, Ashby, Capak, Cooray, de~Putter, Eifler, Flagey, Gong, Habib, Heitmann, Hirata, Jeong, Katti, Korngut, Krause, Lee, Masters, Mauskopf, Melnick, Mennesson, Nguyen, Öberg, Pullen, Raccanelli, Smith, Song, Tolls, Unwin, Venumadhav, Viero, Werner, \& Zemcov}]{Doré2015}
Doré, O., Bock, J., Ashby, M., {et~al.} 2015, Cosmology with the SPHEREX All-Sky Spectral Survey.
\newblock \doarXiv{1412.4872}

\bibitem[{{Faisst} {et~al.}(2019){Faisst}, {Prakash}, {Capak}, \& {Lee}}]{Faisst2019}
{Faisst}, A.~L., {Prakash}, A., {Capak}, P.~L., \& {Lee}, B. 2019, \apjl, 881, L9, \dodoi{10.3847/2041-8213/ab3581}

\bibitem[{{Finlator} \& {Dav{\'e}}(2008)}]{Finlator2008}
{Finlator}, K., \& {Dav{\'e}}, R. 2008, \mnras, 385, 2181, \dodoi{10.1111/j.1365-2966.2008.12991.x}

\bibitem[{{Fletcher} {et~al.}(2019){Fletcher}, {Tang}, {Robertson}, {Nakajima}, {Ellis}, {Stark}, \& {Inoue}}]{Fletcher2019}
{Fletcher}, T.~J., {Tang}, M., {Robertson}, B.~E., {et~al.} 2019, \apj, 878, 87, \dodoi{10.3847/1538-4357/ab2045}

\bibitem[{{Galametz} {et~al.}(2013){Galametz}, {Grazian}, {Fontana}, {Ferguson}, {Ashby}, {Barro}, {Castellano}, {Dahlen}, {Donley}, {Faber}, {Grogin}, {Guo}, {Huang}, {Kocevski}, {Koekemoer}, {Lee}, {McGrath}, {Peth}, {Willner}, {Almaini}, {Cooper}, {Cooray}, {Conselice}, {Dickinson}, {Dunlop}, {Fazio}, {Foucaud}, {Gardner}, {Giavalisco}, {Hathi}, {Hartley}, {Koo}, {Lai}, {de Mello}, {McLure}, {Lucas}, {Paris}, {Pentericci}, {Santini}, {Simpson}, {Sommariva}, {Targett}, {Weiner}, {Wuyts}, \& {CANDELS Team}}]{Galametz2013}
{Galametz}, A., {Grazian}, A., {Fontana}, A., {et~al.} 2013, \apjs, 206, 10, \dodoi{10.1088/0067-0049/206/2/10}

\bibitem[{Geach(2012)}]{Geach2012}
Geach, J.~E. 2012, Monthly Notices of the Royal Astronomical Society, 419, 2633, \dodoi{10.1111/j.1365-2966.2011.19913.x}

\bibitem[{Grogin {et~al.}(2011)Grogin, Kocevski, Faber, Ferguson, Koekemoer, Riess, Acquaviva, Alexander, Almaini, Ashby, {et~al.}}]{Grogin2011}
Grogin, N.~A., Kocevski, D.~D., Faber, S., {et~al.} 2011, The Astrophysical Journal Supplement Series, 197, 35

\bibitem[{Groves {et~al.}(2006)Groves, Heckman, \& Kauffmann}]{Groves2006}
Groves, B.~A., Heckman, T.~M., \& Kauffmann, G. 2006, Monthly Notices of the Royal Astronomical Society, 371, 1559, \dodoi{10.1111/j.1365-2966.2006.10812.x}

\bibitem[{{Guo} {et~al.}(2013){Guo}, {Ferguson}, {Giavalisco}, {Barro}, {Willner}, {Ashby}, {Dahlen}, {Donley}, {Faber}, {Fontana}, {Galametz}, {Grazian}, {Huang}, {Kocevski}, {Koekemoer}, {Koo}, {McGrath}, {Peth}, {Salvato}, {Wuyts}, {Castellano}, {Cooray}, {Dickinson}, {Dunlop}, {Fazio}, {Gardner}, {Gawiser}, {Grogin}, {Hathi}, {Hsu}, {Lee}, {Lucas}, {Mobasher}, {Nandra}, {Newman}, \& {van der Wel}}]{Guo2013}
{Guo}, Y., {Ferguson}, H.~C., {Giavalisco}, M., {et~al.} 2013, \apjs, 207, 24, \dodoi{10.1088/0067-0049/207/2/24}

\bibitem[{Hambleton {et~al.}(2022)Hambleton, Bianco, Street, Bell, Buckley, Graham, Hernitschek, Lund, Mason, Pepper, Prsa, Rabus, Raiteri, Szabo, Szkody, Andreoni, Antoniucci, Balmaverde, Bellm, Bonito, Bono, Botticella, Brocato, Bricman, Cappellaro, Carnerero, Chornock, Clarke, Cowperthwaite, Cucchiara, D'Ammando, Dage, Dall'Ora, Davenport, de~Martino, de~Somma, Criscienzo, Stefano, Drout, Fabrizio, Fiorentino, Gandhi, Garofalo, Giannini, Gomboc, Greggio, Hartigan, Hundertmark, Johnson, Johnson, Jurkic, Khakpash, Leccia, Li, Magurno, Malanchev, Marconi, Margutti, Marinoni, Mauron, Molinaro, Moller, Moniez, Muraveva, Musella, Ngeow, Pastorello, Petrecca, Piranomonte, Ragosta, Reguitti, Righi, Ripepi, Sandoval, Stassun, Stroh, Terreran, Trimble, Tsapras, van Velzen, Venuti, \& Vink}]{Rubin2022}
Hambleton, K.~M., Bianco, F.~B., Street, R., {et~al.} 2022, Rubin Observatory LSST Transients and Variable Stars Roadmap.
\newblock \doarXiv{2208.04499}

\bibitem[{{Hemmati} {et~al.}(2019){Hemmati}, {Capak}, {Pourrahmani}, {Nayyeri}, {Stern}, {Mobasher}, {Darvish}, {Davidzon}, {Ilbert}, {Masters}, \& {Shahidi}}]{Hemmati2019b}
{Hemmati}, S., {Capak}, P., {Pourrahmani}, M., {et~al.} 2019, \apjl, 881, L14, \dodoi{10.3847/2041-8213/ab3418}

\bibitem[{Hemmati {et~al.}(2019)Hemmati, Capak, Masters, Davidzon, Dorè, Kruk, Mobasher, Rhodes, Scolnic, \& Stern}]{Hemmati2019}
Hemmati, S., Capak, P., Masters, D., {et~al.} 2019, The Astrophysical Journal, 877, 117, \dodoi{10.3847/1538-4357/ab1be5}

\bibitem[{{Huertas-Company} \& {Lanusse}(2023)}]{Huertas2023}
{Huertas-Company}, M., \& {Lanusse}, F. 2023, \pasa, 40, e001, \dodoi{10.1017/pasa.2022.55}

\bibitem[{Jafariyazani {et~al.}(2024)Jafariyazani, Masters, Faisst, Teplitz, \& Ilbert}]{Jafariyazani2024}
Jafariyazani, M., Masters, D., Faisst, A.~L., Teplitz, H.~I., \& Ilbert, O. 2024, The Astrophysical Journal, 967, 60, \dodoi{10.3847/1538-4357/ad38b8}

\bibitem[{Kauffmann {et~al.}(2003)Kauffmann, Heckman, Tremonti, Brinchmann, Charlot, White, Ridgway, Brinkmann, Fukugita, Hall, Ivezić, Richards, \& Schneider}]{Kauffmann2003}
Kauffmann, G., Heckman, T.~M., Tremonti, C., {et~al.} 2003, Monthly Notices of the Royal Astronomical Society, 346, 1055–1077, \dodoi{10.1111/j.1365-2966.2003.07154.x}

\bibitem[{Kewley {et~al.}(2001)Kewley, Dopita, Sutherland, Heisler, \& Trevena}]{Kewley2001}
Kewley, L., Dopita, M., Sutherland, R., Heisler, C., \& Trevena, J. 2001, The Astrophysical Journal, 556, \dodoi{10.1086/321545}

\bibitem[{{Kewley} \& {Dopita}(2002)}]{Kewley2002}
{Kewley}, L.~J., \& {Dopita}, M.~A. 2002, \apjs, 142, 35, \dodoi{10.1086/341326}

\bibitem[{{Kewley} {et~al.}(2004){Kewley}, {Geller}, \& {Jansen}}]{Kewley2004}
{Kewley}, L.~J., {Geller}, M.~J., \& {Jansen}, R.~A. 2004, \aj, 127, 2002, \dodoi{10.1086/382723}

\bibitem[{{Kim} \& {Brunner}(2017)}]{stargalaxy2017}
{Kim}, E.~J., \& {Brunner}, R.~J. 2017, \mnras, 464, 4463, \dodoi{10.1093/mnras/stw2672}

\bibitem[{{Koekemoer} {et~al.}(2011){Koekemoer}, {Faber}, {Ferguson}, {Grogin}, {Kocevski}, {Koo}, {Lai}, {Lotz}, {Lucas}, {McGrath}, {Ogaz}, {Rajan}, {Riess}, {Rodney}, {Strolger}, {Casertano}, {Castellano}, {Dahlen}, {Dickinson}, {Dolch}, {Fontana}, {Giavalisco}, {Grazian}, {Guo}, {Hathi}, {Huang}, {van der Wel}, {Yan}, {Acquaviva}, {Alexander}, {Almaini}, {Ashby}, {Barden}, {Bell}, {Bournaud}, {Brown}, {Caputi}, {Cassata}, {Challis}, {Chary}, {Cheung}, {Cirasuolo}, {Conselice}, {Roshan Cooray}, {Croton}, {Daddi}, {Dav{\'e}}, {de Mello}, {de Ravel}, {Dekel}, {Donley}, {Dunlop}, {Dutton}, {Elbaz}, {Fazio}, {Filippenko}, {Finkelstein}, {Frazer}, {Gardner}, {Garnavich}, {Gawiser}, {Gruetzbauch}, {Hartley}, {H{\"a}ussler}, {Herrington}, {Hopkins}, {Huang}, {Jha}, {Johnson}, {Kartaltepe}, {Khostovan}, {Kirshner}, {Lani}, {Lee}, {Li}, {Madau}, {McCarthy}, {McIntosh}, {McLure}, {McPartland}, {Mobasher}, {Moreira}, {Mortlock}, {Moustakas}, {Mozena}, {Nandra}, {Newman}, {Nielsen}, {Niemi}, {Noeske}, {Papovich},
  {Pentericci}, {Pope}, {Primack}, {Ravindranath}, {Reddy}, {Renzini}, {Rix}, {Robaina}, {Rosario}, {Rosati}, {Salimbeni}, {Scarlata}, {Siana}, {Simard}, {Smidt}, {Snyder}, {Somerville}, {Spinrad}, {Straughn}, {Telford}, {Teplitz}, {Trump}, {Vargas}, {Villforth}, {Wagner}, {Wandro}, {Wechsler}, {Weiner}, {Wiklind}, {Wild}, {Wilson}, {Wuyts}, \& {Yun}}]{Koekemoer2011}
{Koekemoer}, A.~M., {Faber}, S.~M., {Ferguson}, H.~C., {et~al.} 2011, \apjs, 197, 36, \dodoi{10.1088/0067-0049/197/2/36}

\bibitem[{Kohonen(1982)}]{Kohonen1982}
Kohonen, T. 1982, Biological Cybernetics, \dodoi{10.1007/BF00337288}

\bibitem[{{Kriek} {et~al.}(2015){Kriek}, {Shapley}, {Reddy}, {Siana}, {Coil}, {Mobasher}, {Freeman}, {de Groot}, {Price}, {Sanders}, {Shivaei}, {Brammer}, {Momcheva}, {Skelton}, {van Dokkum}, {Whitaker}, {Aird}, {Azadi}, {Kassis}, {Bullock}, {Conroy}, {Dav{\'e}}, {Kere{\v{s}}}, \& {Krumholz}}]{Kriek2015}
{Kriek}, M., {Shapley}, A.~E., {Reddy}, N.~A., {et~al.} 2015, \apjs, 218, 15, \dodoi{10.1088/0067-0049/218/2/15}

\bibitem[{{La Torre} {et~al.}(2024){La Torre}, {Sajina}, {Goulding}, {Marchesini}, {Bezanson}, {Pearl}, \& {Sodr{\'e}}}]{Torre2024}
{La Torre}, V., {Sajina}, A., {Goulding}, A.~D., {et~al.} 2024, \aj, 167, 261, \dodoi{10.3847/1538-3881/ad3821}

\bibitem[{Laidler {et~al.}(2007)Laidler, Papovich, Grogin, Idzi, Dickinson, Ferguson, Hilbert, Clubb, \& Ravindranath}]{Laidler2007}
Laidler, V.~G., Papovich, C., Grogin, N.~A., {et~al.} 2007, Publications of the Astronomical Society of the Pacific, 119, 1325, \dodoi{10.1086/523898}

\bibitem[{{Laigle} {et~al.}(2019){Laigle}, {Davidzon}, {Ilbert}, {Devriendt}, {Kashino}, {Pichon}, {Capak}, {Arnouts}, {de la Torre}, {Dubois}, {Gozaliasl}, {Le Borgne}, {Lilly}, {McCracken}, {Salvato}, \& {Slyz}}]{Laigle2019}
{Laigle}, C., {Davidzon}, I., {Ilbert}, O., {et~al.} 2019, \mnras, 486, 5104, \dodoi{10.1093/mnras/stz1054}

\bibitem[{{Lara-L{\'o}pez} {et~al.}(2010){Lara-L{\'o}pez}, {Cepa}, {Bongiovanni}, {P{\'e}rez Garc{\'\i}a}, {Ederoclite}, {Casta{\~n}eda}, {Fern{\'a}ndez Lorenzo}, {Povi{\'c}}, \& {S{\'a}nchez-Portal}}]{Lara2010}
{Lara-L{\'o}pez}, M.~A., {Cepa}, J., {Bongiovanni}, A., {et~al.} 2010, \aap, 521, L53, \dodoi{10.1051/0004-6361/201014803}

\bibitem[{{Li} {et~al.}(2024){Li}, {Grasha}, {Krumholz}, {Wisnioski}, {Sutherland}, {Kewley}, {Chen}, \& {Li}}]{Li2024}
{Li}, S.-L., {Grasha}, K., {Krumholz}, M.~R., {et~al.} 2024, \mnras, 529, 4993, \dodoi{10.1093/mnras/stae869}

\bibitem[{Lovell {et~al.}(2019)Lovell, Acquaviva, Thomas, Iyer, Gawiser, \& Wilkins}]{Lovell2019}
Lovell, C.~C., Acquaviva, V., Thomas, P.~A., {et~al.} 2019, Monthly Notices of the Royal Astronomical Society, 490, 5503, \dodoi{10.1093/mnras/stz2851}

\bibitem[{{Maiolino} \& {Mannucci}(2019)}]{Maiolino2019}
{Maiolino}, R., \& {Mannucci}, F. 2019, \aapr, 27, 3, \dodoi{10.1007/s00159-018-0112-2}

\bibitem[{{Mannucci} {et~al.}(2010){Mannucci}, {Cresci}, {Maiolino}, {Marconi}, \& {Gnerucci}}]{Mannucci2010}
{Mannucci}, F., {Cresci}, G., {Maiolino}, R., {Marconi}, A., \& {Gnerucci}, A. 2010, \mnras, 408, 2115, \dodoi{10.1111/j.1365-2966.2010.17291.x}

\bibitem[{Masters {et~al.}(2015)Masters, Capak, Stern, Ilbert, Salvato, Schmidt, Longo, Rhodes, Paltani, Mobasher, Hoekstra, Hildebrandt, Coupon, Steinhardt, Speagle, Faisst, Kalinich, Brodwin, Brescia, \& Cavuoti}]{Masters2015}
Masters, D., Capak, P., Stern, D., {et~al.} 2015, The Astrophysical Journal, 813, 53, \dodoi{10.1088/0004-637X/813/1/53}

\bibitem[{{Masters} {et~al.}(2019){Masters}, {Stern}, {Cohen}, {Capak}, {Stanford}, {Hernitschek}, {Galametz}, {Davidzon}, {Rhodes}, {Sanders}, {Mobasher}, {Castander}, {Pruett}, \& {Fotopoulou}}]{Masters2019}
{Masters}, D.~C., {Stern}, D.~K., {Cohen}, J.~G., {et~al.} 2019, \apj, 877, 81, \dodoi{10.3847/1538-4357/ab184d}

\bibitem[{{McCullough} {et~al.}(2024){McCullough}, {Gruen}, {Amon}, {Roodman}, {Masters}, {Raichoor}, {Schlegel}, {Canning}, {Castander}, {DeRose}, {Miquel}, {Myles}, {Newman}, {Slosar}, {Speagle}, {Wilson}, {Aguilar}, {Ahlen}, {Bailey}, {Brooks}, {Claybaugh}, {Cole}, {Dawson}, {de la Macorra}, {Doel}, {Forero-Romero}, {Gontcho A Gontcho}, {Guy}, {Kehoe}, {Kremin}, {Landriau}, {Le Guillou}, {Levi}, {Manera}, {Martini}, {Meisner}, {Moustakas}, {Nie}, {Percival}, {Poppett}, {Prada}, {Rezaie}, {Rossi}, {Sanchez}, {Seo}, {Tarl{\'e}}, {Weaver}, {Zhou}, {Zou}, \& {DESI Collaboration}}]{McCullough2024}
{McCullough}, J., {Gruen}, D., {Amon}, A., {et~al.} 2024, \mnras, 531, 2582, \dodoi{10.1093/mnras/stae1316}

\bibitem[{{McInnes} {et~al.}(2018){McInnes}, {Healy}, \& {Melville}}]{McInnes2018}
{McInnes}, L., {Healy}, J., \& {Melville}, J. 2018, arXiv e-prints, arXiv:1802.03426, \dodoi{10.48550/arXiv.1802.03426}

\bibitem[{McLean {et~al.}(2012)McLean, Berger, \& Reiners}]{McLean2012}
McLean, M., Berger, E., \& Reiners, A. 2012, The Astrophysical Journal, 746, 23, \dodoi{10.1088/0004-637X/746/1/23}

\bibitem[{Moosavi {et~al.}(2014)Moosavi, Packmann, \& Vall{\'e}s}]{sompy}
Moosavi, V., Packmann, S., \& Vall{\'e}s, I. 2014, SOMPY: A Python Library for Self Organizing Map (SOM)

\bibitem[{{Nagao} {et~al.}(2006){Nagao}, {Maiolino}, \& {Marconi}}]{Nagao2006}
{Nagao}, T., {Maiolino}, R., \& {Marconi}, A. 2006, \aap, 459, 85, \dodoi{10.1051/0004-6361:20065216}

\bibitem[{{Nakajima} {et~al.}(2016){Nakajima}, {Ellis}, {Iwata}, {Inoue}, {Kusakabe}, {Ouchi}, \& {Robertson}}]{Nakajima2016}
{Nakajima}, K., {Ellis}, R.~S., {Iwata}, I., {et~al.} 2016, \apjl, 831, L9, \dodoi{10.3847/2041-8205/831/1/L9}

\bibitem[{{Nakajima} {et~al.}(2020){Nakajima}, {Ellis}, {Robertson}, {Tang}, \& {Stark}}]{Nakajima2020}
{Nakajima}, K., {Ellis}, R.~S., {Robertson}, B.~E., {Tang}, M., \& {Stark}, D.~P. 2020, \apj, 889, 161, \dodoi{10.3847/1538-4357/ab6604}

\bibitem[{{Nakajima} {et~al.}(2023){Nakajima}, {Ouchi}, {Isobe}, {Harikane}, {Zhang}, {Ono}, {Umeda}, \& {Oguri}}]{Nakajima2023}
{Nakajima}, K., {Ouchi}, M., {Isobe}, Y., {et~al.} 2023, \apjs, 269, 33, \dodoi{10.3847/1538-4365/acd556}

\bibitem[{Nayyeri {et~al.}(2017)Nayyeri, Hemmati, Mobasher, Ferguson, Cooray, Barro, Faber, Dickinson, Koekemoer, Peth, Salvato, Ashby, Darvish, Donley, Durbin, Finkelstein, Fontana, Grogin, Gruetzbauch, Huang, Khostovan, Kocevski, Kodra, Lee, Newman, Pacifici, Pforr, Stefanon, Wiklind, Willner, Wuyts, Castellano, Conselice, Dolch, Dunlop, Galametz, Hathi, Lucas, \& Yan}]{Nayyeri2017}
Nayyeri, H., Hemmati, S., Mobasher, B., {et~al.} 2017, The Astrophysical Journal Supplement Series, 228, 7, \dodoi{10.3847/1538-4365/228/1/7}

\bibitem[{{Noeske} {et~al.}(2007){Noeske}, {Weiner}, {Faber}, {Papovich}, {Koo}, {Somerville}, {Bundy}, {Conselice}, {Newman}, {Schiminovich}, {Le Floc'h}, {Coil}, {Rieke}, {Lotz}, {Primack}, {Barmby}, {Cooper}, {Davis}, {Ellis}, {Fazio}, {Guhathakurta}, {Huang}, {Kassin}, {Martin}, {Phillips}, {Rich}, {Small}, {Willmer}, \& {Wilson}}]{Noeske2007}
{Noeske}, K.~G., {Weiner}, B.~J., {Faber}, S.~M., {et~al.} 2007, \apjl, 660, L43, \dodoi{10.1086/517926}

\bibitem[{{Oke} \& {Gunn}(1983)}]{Oke1983}
{Oke}, J.~B., \& {Gunn}, J.~E. 1983, \apj, 266, 713, \dodoi{10.1086/160817}

\bibitem[{{Parker} {et~al.}(2024){Parker}, {Lanusse}, {Golkar}, {Sarra}, {Cranmer}, {Bietti}, {Eickenberg}, {Krawezik}, {McCabe}, {Morel}, {Ohana}, {Pettee}, {R{\'e}galdo-Saint Blancard}, {Cho}, {Ho}, \& {Polymathic AI Collaboration}}]{Parker2024}
{Parker}, L., {Lanusse}, F., {Golkar}, S., {et~al.} 2024, \mnras, 531, 4990, \dodoi{10.1093/mnras/stae1450}

\bibitem[{{Pourrahmani} {et~al.}(2018){Pourrahmani}, {Nayyeri}, \& {Cooray}}]{Lensfind2018}
{Pourrahmani}, M., {Nayyeri}, H., \& {Cooray}, A. 2018, \apj, 856, 68, \dodoi{10.3847/1538-4357/aaae6a}

\bibitem[{Racca {et~al.}(2016)Racca, Laureijs, Stagnaro, Salvignol, Lorenzo~Alvarez, Saavedra~Criado, Gaspar~Venancio, Short, Strada, Bönke, Colombo, Calvi, Maiorano, Piersanti, Prezelus, Rosato, Pinel, Rozemeijer, Lesna, Musi, Sias, Anselmi, Cazaubiel, Vaillon, Mellier, Amiaux, Berthé, Sauvage, Azzollini, Cropper, Pottinger, Jahnke, Ealet, Maciaszek, Pasian, Zacchei, Scaramella, Hoar, Kohley, Vavrek, Rudolph, \& Schmidt}]{Racca2016}
Racca, G.~D., Laureijs, R., Stagnaro, L., {et~al.} 2016, in Space Telescopes and Instrumentation 2016: Optical, Infrared, and Millimeter Wave, ed. H.~A. MacEwen, G.~G. Fazio, M.~Lystrup, N.~Batalha, N.~Siegler, \& E.~C. Tong (SPIE), \dodoi{10.1117/12.2230762}

\bibitem[{Raouf {et~al.}(2016)Raouf, Khosroshahi, \& Dariush}]{Raouf2016}
Raouf, M., Khosroshahi, H.~G., \& Dariush, A. 2016, The Astrophysical Journal, 824, 140, \dodoi{10.3847/0004-637x/824/2/140}

\bibitem[{Reddy {et~al.}(2015)Reddy, Kriek, Shapley, Freeman, Siana, Coil, Mobasher, Price, Sanders, \& Shivaei}]{Reddy2015}
Reddy, N.~A., Kriek, M., Shapley, A.~E., {et~al.} 2015, The Astrophysical Journal, 806, 259, \dodoi{10.1088/0004-637X/806/2/259}

\bibitem[{{Runco} {et~al.}(2022){Runco}, {Reddy}, {Shapley}, {Steidel}, {Sanders}, {Strom}, {Coil}, {Kriek}, {Mobasher}, {Pettini}, {Rudie}, {Siana}, {Topping}, {Trainor}, {Freeman}, {Shivaei}, {Azadi}, {Price}, {Leung}, {Fetherolf}, {de Groot}, {Zick}, {Fornasini}, \& {Barro}}]{Runco2022}
{Runco}, J.~N., {Reddy}, N.~A., {Shapley}, A.~E., {et~al.} 2022, \mnras, 513, 3871, \dodoi{10.1093/mnras/stac1115}

\bibitem[{Sanders {et~al.}(2021)Sanders, Shapley, Jones, Reddy, Kriek, Siana, Coil, Mobasher, Shivaei, Davé, Azadi, Price, Leung, Freeman, Fetherolf, de~Groot, Zick, \& Barro}]{Sanders2021}
Sanders, R.~L., Shapley, A.~E., Jones, T., {et~al.} 2021, The Astrophysical Journal, 914, 19, \dodoi{10.3847/1538-4357/abf4c1}

\bibitem[{Scoville {et~al.}(2007)Scoville, Aussel, Benson, Blain, Calzetti, Capak, Ellis, El-Zant, Finoguenov, Giavalisco, {et~al.}}]{Scoville2007}
Scoville, N., Aussel, H., Benson, A., {et~al.} 2007, The Astrophysical Journal Supplement Series, 172, 150

\bibitem[{{Shivaei} {et~al.}(2015){Shivaei}, {Reddy}, {Shapley}, {Kriek}, {Siana}, {Mobasher}, {Coil}, {Freeman}, {Sanders}, {Price}, {de Groot}, \& {Azadi}}]{Shivaei2015}
{Shivaei}, I., {Reddy}, N.~A., {Shapley}, A.~E., {et~al.} 2015, \apj, 815, 98, \dodoi{10.1088/0004-637X/815/2/98}

\bibitem[{{Speagle} {et~al.}(2014){Speagle}, {Steinhardt}, {Capak}, \& {Silverman}}]{Speakle2014}
{Speagle}, J.~S., {Steinhardt}, C.~L., {Capak}, P.~L., \& {Silverman}, J.~D. 2014, \apjs, 214, 15, \dodoi{10.1088/0067-0049/214/2/15}

\bibitem[{{Stefanon} {et~al.}(2017){Stefanon}, {Yan}, {Mobasher}, {Barro}, {Donley}, {Fontana}, {Hemmati}, {Koekemoer}, {Lee}, {Lee}, {Nayyeri}, {Peth}, {Pforr}, {Salvato}, {Wiklind}, {Wuyts}, {Ashby}, {Castellano}, {Conselice}, {Cooper}, {Cooray}, {Dolch}, {Ferguson}, {Galametz}, {Giavalisco}, {Guo}, {Willner}, {Dickinson}, {Faber}, {Fazio}, {Gardner}, {Gawiser}, {Grazian}, {Grogin}, {Kocevski}, {Koo}, {Lee}, {Lucas}, {McGrath}, {Nandra}, {Newman}, \& {van der Wel}}]{Stefanon2017}
{Stefanon}, M., {Yan}, H., {Mobasher}, B., {et~al.} 2017, \apjs, 229, 32, \dodoi{10.3847/1538-4365/aa66cb}

\bibitem[{{Stern} {et~al.}(2005){Stern}, {Eisenhardt}, {Gorjian}, {Kochanek}, {Caldwell}, {Eisenstein}, {Brodwin}, {Brown}, {Cool}, {Dey}, {Green}, {Jannuzi}, {Murray}, {Pahre}, \& {Willner}}]{Stern2005}
{Stern}, D., {Eisenhardt}, P., {Gorjian}, V., {et~al.} 2005, \apj, 631, 163, \dodoi{10.1086/432523}

\bibitem[{{Strom} {et~al.}(2017){Strom}, {Steidel}, {Rudie}, {Trainor}, {Pettini}, \& {Reddy}}]{strom2017}
{Strom}, A.~L., {Steidel}, C.~C., {Rudie}, G.~C., {et~al.} 2017, \apj, 836, 164, \dodoi{10.3847/1538-4357/836/2/164}

\bibitem[{Surana {et~al.}(2020)Surana, Wadadekar, Bait, \& Bhosale}]{Surana2020}
Surana, S., Wadadekar, Y., Bait, O., \& Bhosale, H. 2020, Monthly Notices of the Royal Astronomical Society, 493, 4808, \dodoi{10.1093/mnras/staa537}

\bibitem[{Sánchez~Almeida {et~al.}(2018)Sánchez~Almeida, Caon, Muñoz-Tuñón, Filho, \& Cerviño}]{Almeida2018}
Sánchez~Almeida, J., Caon, N., Muñoz-Tuñón, C., Filho, M., \& Cerviño, M. 2018, Monthly Notices of the Royal Astronomical Society, 476, 4765, \dodoi{10.1093/mnras/sty510}

\bibitem[{Tamura {et~al.}(2016)Tamura, Takato, Shimono, Moritani, Yabe, Ishizuka, Ueda, Kamata, Aghazarian, Arnouts, Barban, Barkhouser, Borges, Braun, Carr, Chabaud, Chang, Chen, Chiba, Chou, Chu, Cohen, de~Almeida, de~Oliveira, de~Oliveira, Dekany, Dohlen, dos Santos, dos Santos, Ellis, Fabricius, Ferrand, Ferreira, Golebiowski, Greene, Gross, Gunn, Hammond, Harding, Hart, Heckman, Hirata, Ho, Hope, Hovland, Hsu, Hu, Huang, Jaquet, Jing, Karr, Kimura, King, Komatsu, Le~Brun, Le~Fèvre, Le~Fur, Le~Mignant, Ling, Loomis, Lupton, Madec, Mao, Marrara, Mendes~de Oliveira, Minowa, Morantz, Murayama, Murray, Ohyama, Orndorff, Pascal, Pereira, Reiley, Reinecke, Ritter, Roberts, Schwochert, Seiffert, Smee, Sodre, Spergel, Steinkraus, Strauss, Surace, Suto, Suzuki, Swinbank, Tait, Takada, Tamura, Tanaka, Tresse, Verducci, Vibert, Vidal, Wang, Wen, Yan, \& Yasuda}]{Tamura2016}
Tamura, N., Takato, N., Shimono, A., {et~al.} 2016, in Ground-based and Airborne Instrumentation for Astronomy VI, ed. C.~J. Evans, L.~Simard, \& H.~Takami (SPIE), \dodoi{10.1117/12.2232103}

\bibitem[{{Trainor} {et~al.}(2016){Trainor}, {Strom}, {Steidel}, \& {Rudie}}]{Trainor2016}
{Trainor}, R.~F., {Strom}, A.~L., {Steidel}, C.~C., \& {Rudie}, G.~C. 2016, \apj, 832, 171, \dodoi{10.3847/0004-637X/832/2/171}

\bibitem[{{Tremonti} {et~al.}(2004){Tremonti}, {Heckman}, {Kauffmann}, {Brinchmann}, {Charlot}, {White}, {Seibert}, {Peng}, {Schlegel}, {Uomoto}, {Fukugita}, \& {Brinkmann}}]{Tremonti2004}
{Tremonti}, C.~A., {Heckman}, T.~M., {Kauffmann}, G., {et~al.} 2004, \apj, 613, 898, \dodoi{10.1086/423264}

\bibitem[{Trump {et~al.}(2015)Trump, Sun, Zeimann, Luck, Bridge, Grier, Hagen, Juneau, Montero-Dorta, Rosario, Brandt, Ciardullo, \& Schneider}]{Trump2015}
Trump, J.~R., Sun, M., Zeimann, G.~R., {et~al.} 2015, The Astrophysical Journal, 811, 26, \dodoi{10.1088/0004-637X/811/1/26}

\bibitem[{{van den Busch} {et~al.}(2022){van den Busch}, {Wright}, {Hildebrandt}, {Bilicki}, {Asgari}, {Joudaki}, {Blake}, {Heymans}, {Kannawadi}, {Shan}, \& {Tr{\"o}ster}}]{Busch2022}
{van den Busch}, J.~L., {Wright}, A.~H., {Hildebrandt}, H., {et~al.} 2022, \aap, 664, A170, \dodoi{10.1051/0004-6361/202142083}

\bibitem[{Villaescusa-Navarro {et~al.}(2021)Villaescusa-Navarro, Anglés-Alcázar, Genel, Spergel, S.~Somerville, Dave, Pillepich, Hernquist, Nelson, Torrey, Narayanan, Li, Philcox, La~Torre, Maria~Delgado, Ho, Hassan, Burkhart, Wadekar, Battaglia, Contardo, \& Bryan}]{Villaescusa_Navarro2021}
Villaescusa-Navarro, F., Anglés-Alcázar, D., Genel, S., {et~al.} 2021, The Astrophysical Journal, 915, 71, \dodoi{10.3847/1538-4357/abf7ba}

\bibitem[{Villaescusa-Navarro {et~al.}(2022)Villaescusa-Navarro, Ding, Genel, Tonnesen, La~Torre, Spergel, Teyssier, Li, Heneka, Lemos, Anglés-Alcázar, Nagai, \& Vogelsberger}]{Villaescusa_Navarro2022}
Villaescusa-Navarro, F., Ding, J., Genel, S., {et~al.} 2022, The Astrophysical Journal, 929, 132, \dodoi{10.3847/1538-4357/ac5d3f}

\bibitem[{Xue {et~al.}(2010)Xue, Brandt, Luo, Rafferty, Alexander, Bauer, Lehmer, Schneider, \& Silverman}]{Xue2010}
Xue, Y.~Q., Brandt, W.~N., Luo, B., {et~al.} 2010, The Astrophysical Journal, 720, 368–391, \dodoi{10.1088/0004-637x/720/1/368}

\bibitem[{{York} {et~al.}(2000){York}, {Adelman}, {Anderson}, {Anderson}, {Annis}, {Bahcall}, {Bakken}, {Barkhouser}, {Bastian}, {Berman}, {Boroski}, {Bracker}, {Briegel}, {Briggs}, {Brinkmann}, {Brunner}, {Burles}, {Carey}, {Carr}, {Castander}, {Chen}, {Colestock}, {Connolly}, {Crocker}, {Csabai}, {Czarapata}, {Davis}, {Doi}, {Dombeck}, {Eisenstein}, {Ellman}, {Elms}, {Evans}, {Fan}, {Federwitz}, {Fiscelli}, {Friedman}, {Frieman}, {Fukugita}, {Gillespie}, {Gunn}, {Gurbani}, {de Haas}, {Haldeman}, {Harris}, {Hayes}, {Heckman}, {Hennessy}, {Hindsley}, {Holm}, {Holmgren}, {Huang}, {Hull}, {Husby}, {Ichikawa}, {Ichikawa}, {Ivezi{\'c}}, {Kent}, {Kim}, {Kinney}, {Klaene}, {Kleinman}, {Kleinman}, {Knapp}, {Korienek}, {Kron}, {Kunszt}, {Lamb}, {Lee}, {Leger}, {Limmongkol}, {Lindenmeyer}, {Long}, {Loomis}, {Loveday}, {Lucinio}, {Lupton}, {MacKinnon}, {Mannery}, {Mantsch}, {Margon}, {McGehee}, {McKay}, {Meiksin}, {Merelli}, {Monet}, {Munn}, {Narayanan}, {Nash}, {Neilsen}, {Neswold}, {Newberg}, {Nichol}, {Nicinski},
  {Nonino}, {Okada}, {Okamura}, {Ostriker}, {Owen}, {Pauls}, {Peoples}, {Peterson}, {Petravick}, {Pier}, {Pope}, {Pordes}, {Prosapio}, {Rechenmacher}, {Quinn}, {Richards}, {Richmond}, {Rivetta}, {Rockosi}, {Ruthmansdorfer}, {Sandford}, {Schlegel}, {Schneider}, {Sekiguchi}, {Sergey}, {Shimasaku}, {Siegmund}, {Smee}, {Smith}, {Snedden}, {Stone}, {Stoughton}, {Strauss}, {Stubbs}, {SubbaRao}, {Szalay}, {Szapudi}, {Szokoly}, {Thakar}, {Tremonti}, {Tucker}, {Uomoto}, {Vanden Berk}, {Vogeley}, {Waddell}, {Wang}, {Watanabe}, {Weinberg}, {Yanny}, {Yasuda}, \& {SDSS Collaboration}}]{York2000}
{York}, D.~G., {Adelman}, J., {Anderson}, John~E., J., {et~al.} 2000, \aj, 120, 1579, \dodoi{10.1086/301513}

\bibitem[{Zou {et~al.}(2019)Zou, Yang, Brandt, \& Xue}]{Zou2019}
Zou, F., Yang, G., Brandt, W.~N., \& Xue, Y. 2019, The Astrophysical Journal, 878, 11, \dodoi{10.3847/1538-4357/ab1eb1}

\end{thebibliography}

\end{document}